\documentclass[twocolumn]{aastex62}

\graphicspath{{./}{figures/}}

\shorttitle{Coldest CatWISE Brown Dwarfs}
\shortauthors{Meisner et al.}

\usepackage{mathtools}

\begin{document}

\title{Expanding the Y Dwarf Census with \textit{Spitzer} Follow-up of the \\ Coldest CatWISE Solar Neighborhood Discoveries}

\correspondingauthor{Aaron M. Meisner}
\email{ameisner@noao.edu}

\author[0000-0002-1125-7384]{Aaron M. Meisner}
\affiliation{NSF's National Optical-Infrared Astronomy Research Laboratory, 950 N. Cherry Ave., Tucson, AZ 85719, USA}

\author{Dan Caselden}
\affiliation{Gigamon Applied Threat Research, 619 Western Avenue, Suite 200, Seattle, WA 98104, USA}

\author[0000-0003-4269-260X]{J. Davy Kirkpatrick}
\affiliation{IPAC, Mail Code 100-22, California Institute of Technology, 1200 E. California Blvd., Pasadena, CA 91125, USA}

\author[0000-0001-7519-1700]{Federico Marocco}\altaffiliation{NASA Postdoctoral Program Fellow}
\affiliation{Jet Propulsion Laboratory, California Institute of Technology, 4800 Oak Grove Drive, M/S 169-327, Pasadena, CA 91109, USA}
\affiliation{IPAC, Mail Code 100-22, California Institute of Technology, 1200 E. California Blvd., Pasadena, CA 91125, USA}

\author{Christopher R. Gelino}
\affiliation{IPAC, Mail Code 100-22, California Institute of Technology, 1200 E. California Blvd., Pasadena, CA 91125, USA}

\author[0000-0001-7780-3352]{Michael C. Cushing}
\affiliation{Department of Physics and Astronomy, University of Toledo, 2801 West Bancroft St., Toledo, OH 43606, USA}

\author{Peter R. M. Eisenhardt}
\affiliation{Jet Propulsion Laboratory, California Institute of Technology, 4800 Oak Grove Drive, M/S 169-327, Pasadena, CA 91109, USA}

\author[0000-0001-5058-1593]{Edward L. Wright}
\affiliation{Department of Physics and Astronomy, UCLA, 430 Portola Plaza, Box 951547, Los Angeles, CA 90095-1547, USA}

\author[0000-0001-6251-0573]{Jacqueline K. Faherty}
\affiliation{Department of Astrophysics, American Museum of Natural History, Central Park West at 79th Street, NY 10024, USA}

\author{Renata Koontz}
\affiliation{University of California, Riverside, 900 University Ave, Riverside, CA 92521, USA}

\author{Elijah J. Marchese}
\affiliation{University of California, Riverside, 900 University Ave, Riverside, CA 92521, USA}

\author{Mohammed Khalil}
\affiliation{International College, PO Box 113-5373 Hamra, Bliss Street, Beirut, Lebanon}
\affiliation{IPAC, Mail Code 100-22, California Institute of Technology, 1200 E. California Blvd., Pasadena, CA 91125, USA}
\affiliation{Stanford University, 450 Serra Mall, Stanford, CA 94305, USA}

\author{John W. Fowler}
\affiliation{230 Pacific St., Apt. 205, Santa Monica, CA 90405, USA}

\author[0000-0002-3569-7421]{Edward F. Schlafly}
\affiliation{Lawrence Berkeley National Laboratory, Berkeley, CA, 94720, USA}

\begin{abstract}

We present \textit{Spitzer} 3.6$\mu$m and 4.5$\mu$m follow-up of 170 candidate extremely cool brown dwarfs newly discovered via the combination of WISE and NEOWISE imaging at 3$-$5$\mu$m. CatWISE, a joint analysis of archival WISE and NEOWISE data, has improved upon the motion measurements of AllWISE by leveraging a $>$10$\times$ time baseline enhancement, from 0.5 years (AllWISE) to 6.5 years (CatWISE). As a result, CatWISE motion selection has yielded a large sample of previously unrecognized brown dwarf candidates, many of which have archival detections exclusively in the WISE 4.6$\mu$m (W2) channel, suggesting that they could be both exceptionally cold and nearby. Where these objects go undetected in WISE W1 (3.4$\mu$m), \textit{Spitzer} can provide critically informative detections at 3.6$\mu$m. Of our motion-confirmed discoveries, seventeen have a best-fit \textit{Spitzer} [3.6]$-$[4.5] color most consistent with spectral type Y. CWISEP J144606.62$-$231717.8 ($\mu \approx 1.3''$/yr) is likely the reddest, and therefore potentially coldest, member of our sample with a very uncertain [3.6]$-$[4.5] color of 3.71 $\pm$ 0.44 magnitudes. We also highlight our highest proper motion discovery, WISEA J153429.75$-$104303.3, with $\mu \approx 2.7''$/yr. Given that the prior list of confirmed and presumed Y dwarfs consists of just 27 objects, the \textit{Spitzer} follow-up presented in this work has substantially expanded the sample of identified Y dwarfs. Our new discoveries thus represent significant progress toward understanding the bottom of the substellar mass function, investigating the diversity of the Y dwarf population, and selecting optimal brown dwarf targets for JWST spectroscopy.

\end{abstract}


\section{Introduction}
\label{sec:intro}

How complete is our census of the Sun's closest neighbors? How far does the population of substellar objects born like stars extend into the planetary-mass regime? The Wide-field Infrared Survey Explorer \citep[WISE;][]{wright10}, with its unique full-sky sensitivity at 4.6$\mu$m, has unrivaled potential to answer these open questions by identifying the coldest brown dwarfs down to planetary masses. WISE-based discoveries already include the three nearest known brown dwarfs, among which is the coldest known brown dwarf (WISE 0855$-$0714), a $\sim$250 K planetary-mass object \citep{luhman_planetx, j0855, luhman16ab}. \cite{wright_0855} estimate that between 3 and 34 WISE 0855$-$0714 analogs should be detectable with WISE, yet only perhaps one such candidate has thus far been found \citep{marocco2019}.

The coolest brown dwarfs revealed by WISE will be key \textit{James Webb Space Telescope}  \citep[JWST;][]{jwst} targets, overlapping in mass and temperature with extrasolar giant planets and providing simplified laboratories for modeling planetary atmospheres, free of the irradiation and contaminating glare from a primary star. Indeed, WISE has already established the existence of a new brown dwarf spectral class with  $T_{\textrm{eff}} \lesssim 500$ K \citep[Y dwarfs;][]{cushing_y_dwarfs, kirkpatrick11}. Although prior WISE brown dwarf searches have been highly successful \citep[e.g.,][]{cushing_y_dwarfs, kirkpatrick11, kirkpatrick12, griffith_brown_dwarfs, mace_t_dwarfs, luhman_planetx, allwise_motion_survey, allwise2_motion_survey, pinfield_methodology, schneider_neowise, backyard_worlds, tinney18, burningham_review}, the Y dwarf census has stagnated in recent years at a sample size of just $\sim$25-30 objects.

The discoveries of nearly all known Y dwarfs can be traced back to WISE. The WISE 3.4$\mu$m (W1) and 4.6$\mu$m (W2) bands were designed for optimal sensitivity to the coldest brown dwarfs \citep{mainzer_first_brown_dwarf}, such that selecting for large W1$-$W2 color is a highly effective search strategy \citep[e.g.,][]{griffith_brown_dwarfs, kirkpatrick11, kirkpatrick12}. Even more powerful is the combination of WISE-based color and motion criteria, eliminating stationary extragalactic contaminants \citep[e.g.,][]{luhman_planetx, j0855}. \cite{pinfield_y_dwarf} illustrated that faint, unrecognized Y dwarfs remain to be found in the WISE imaging, identifiable with novel search criteria using WISE data alone. Follow-up techniques such as methane on/off imaging \citep{tinney_methodology, tinney18} and adaptive optics imaging \citep{liu_binary_j1217, dupuy_binary_j0146} have also proven effective at pinpointing Y dwarfs among the numerous WISE solar neighborhood discoveries.

Our ``CatWISE'' analysis \citep{catwise_data_paper} represents a major step toward realizing the entire WISE data set's full sensitivity for brown dwarf discovery.
Mining the vast WISE imaging archive to its faintest depths is a formidable challenge, and prior WISE motion searches have generally been limited by restricting to bright single-exposure detections and/or the short half-year time baseline of 2010-2011 observations. CatWISE pushes several magnitudes deeper than foregoing WISE-based motion surveys by jointly analyzing four years of WISE and NEOWISE \citep{neowise, neowiser} data spanning the 2010-2016 time period. CatWISE thereby provides long time baseline WISE proper motions for roughly a billion mid-infrared sources over the full sky.

Using the CatWISE Preliminary catalog\footnote{\url{https://catwise.github.io}} and drawing upon well-established faint moving object selection/confirmation techniques \citep[e.g.,][]{superblink}, we have performed an extensive motion-based search for previously undiscovered solar neighborhood constituents. As part of our ground and space based follow-up program, we have obtained \textit{Spitzer} IRAC \citep{spitzer_overview, irac} photometry of $\sim$170 newly discovered brown dwarf candidates suspected of being extremely cold and/or nearby. At CatWISE depths, the hallmark of such targets is the presence of a moving 4.6$\mu$m (W2) source with no firmly detected 3.4$\mu$m (W1) counterpart.

Detectable motion and a W2 magnitude alone are insufficient to estimate the basic parameters of most immediate interest for these brown dwarf candidates: spectral type, temperature, luminosity, distance, and near-infrared flux. The mid-infrared color,
whether W1$-$W2 from WISE or [3.6]$-$[4.5] from \textit{Spitzer}, represents a critical diagnostic in obtaining estimates for all of these quantities, as both colors tend to increase monotonically toward later spectral types beyond mid-T \citep[e.g.,][]{patten06, kirkpatrick11}. For our targets with W1 non-detections, \textit{Spitzer} provides the only opportunity to measure a mid-infrared color. With spectral type estimates based on IRAC colors in hand, luminosity, distance and near-infrared flux estimates also follow.  

Based on the \textit{Spitzer} follow-up we present in this work, many of our discoveries have photometric spectral type estimates placing them within the 20 pc volume and/or very red [3.6]$-$[4.5] colors most consistent with spectral type Y. Among these, CWISEP J144606.62$-$231717.8 (hereafter CWISEP 1446$-$2317) stands out with an exceptionally large but highly uncertain [3.6]$-$[4.5] color of 3.71 $\pm$ 0.44 magnitudes.

In $\S$\ref{sec:wise_overview} we briefly summarize relevant characteristics of the WISE and NEOWISE missions. In $\S$\ref{sec:catwise} we provide a concise overview of CatWISE. In $\S$\ref{sec:target_selection} we describe our selection of brown dwarf candidate targets for follow-up \textit{Spitzer} photometry. In $\S$\ref{sec:sample_properties} we present the basic properties of our \textit{Spitzer} photometry target sample. In $\S$\ref{sec:observing_strategy} we explain our \textit{Spitzer} observing strategy. In $\S$\ref{sec:spitzer_photometry} we present our \textit{Spitzer} color measurements. In $\S$\ref{sec:astrometry} we combine WISE and \textit{Spitzer} astrometry to confirm the motions of our brown dwarf candidates. In $\S$\ref{sec:ground_based} we present complementary near-infrared photometry drawn from our ground-based follow-up observations and archival data sets. We discuss the combined implications of our photometric and astrometric analyses in $\S$\ref{sec:discussion}. We conclude in $\S$\ref{sec:conclusion}.

\section{WISE/NEOWISE Overview}
\label{sec:wise_overview}

Launched into low-Earth orbit in late 2009, WISE is a satellite-borne 40 cm aperture telescope. During early and mid 2010, WISE mapped the entire sky in four broad infrared bandpasses centered at 3.4$\mu$m (W1), 4.6$\mu$m (W2), 12$\mu$m (W3) and 22$\mu$m (W4). Although observations in the two longest wavelength channels were discontinued by late 2010 due to cryogen depletion, WISE kept observing in W1 and W2 until 2011 February as part of the NEOWISE mission \citep{neowise}. WISE was placed into hibernation from 2011 February until 2013 December, at which point it recommenced surveying in W1 and W2 thanks to the NEOWISE-Reactivation \citep[NEOWISE-R;][]{neowiser} mission extension. WISE has continued observations since reactivation to this writing (October 2019). A typical sky location is observed during a $\sim$1 day time period once every six months, and WISE has now performed a total of more than 13 complete sky passes in W1 and W2. The high quality of W1 and W2 imaging has remained essentially unchanged throughout the entire WISE lifetime \citep{neowiser, cutri15}.

\section{CatWISE}
\label{sec:catwise}

Although NEOWISE-R has now supplied the vast majority of W1/W2 observations, the mission itself does not provide any coadded data products optimized for science beyond the inner solar system. As a result, AllWISE \citep{cutri13} has remained the definitive coadded WISE catalog for many years despite incorporating only the $\sim$13 months of pre-hibernation WISE imaging.

Our CatWISE archival data analysis program \citep{catwise_data_paper} has combined $\sim$4 years of 2010-2016 WISE/NEOWISE data to build a deeper, longer time baseline successor to AllWISE at 3$-$5$\mu$m. Whereas the AllWISE Source Catalog directly modeled WISE single-exposure images, CatWISE instead applies the AllWISE cataloging software to ``unWISE'' coadds \citep{lang14} as a computational convenience. By detecting W1/W2 sources in four-year depth unWISE stacks \citep{fulldepth_neo3}, CatWISE extracts $5\sigma$ sources to Vega magnitudes of W1 = 17.67 and W2 = 16.47, $\sim$0.6 mag deeper than AllWISE\footnote{Magnitudes and colors quoted throughout this paper are in the Vega system unless otherwise noted.}. CatWISE fits apparent linear motions for every source using a set of ``time-resolved'' unWISE coadds \citep{tr_neo2, tr_neo3}. Each time-resolved coadd stacks the $\sim$1 day of WISE frames together at a given sky location during a single sky pass, sampling the motion at 6 month intervals. Such coaddition results in effectively no loss of motion information for objects in the solar neighborhood\footnote{The time-resolved unWISE coaddition would only begin to incur significant smearing of fast-moving sources at linear motions of order 100$''$/yr. For comparison, the largest proper motion of any known star or brown dwarf is 10.4$''$/yr (Barnard's Star).} ($\mu \lesssim 10''$/yr). By virtue of its $>$10$\times$ extended time baseline and 4$\times$ input imaging increase, CatWISE derives motions an order of magnitude more accurate than those of AllWISE for $\sim$900 million sources over the full sky\footnote{CatWISE can be queried via IRSA at \url{https://irsa.ipac.caltech.edu/applications/Gator/}.}. We should therefore expect CatWISE motion and/or color selections to reveal many nearby brown dwarfs not previously identified with AllWISE. This includes objects below the AllWISE detection limit and also those with AllWISE motion measurements too noisy to be statistically significant.

Artifact flagging is a key ingredient in WISE-based rare object searches, where anomalies due to bright stars and blending dramatically outnumber the astrophysical sources of interest. Our brown dwarf  searches take advantage of two complementary artifact flagging capabilities provided by CatWISE: (1) CC flags inherited from AllWISE via cross-match and (2) unWISE ab\_flags similar to the CC flags, but available even in cases when a CatWISE source has no AllWISE counterpart\footnote{unWISE artifact flagging is described in detail in the appendix of \cite{neo4_coadds}.}.

\section{\textit{Spitzer} Target Selection}
\label{sec:target_selection}

Our \textit{Spitzer} follow-up observations presented throughout this work were acquired as part of program 14034 (hereafter p14034; PI Meisner). At the time of our Cycle 14 proposal submission (2018 March), \textit{Spitzer}'s final observations were expected to take place on 2019 November 30. Given \textit{Spitzer}'s impending retirement, we sought to fill our target list with the most exceptional WISE-based cold brown dwarf candidates selected by any/all means necessary, and as a result virtually no emphasis was placed on sample uniformity/homogeneity.

Additionally, our search for \textit{Spitzer} targets was performed as part of a larger effort to fully mine CatWISE (and unWISE/AllWISE) for moving object discoveries, including earlier type candidates accessible via our ground-based photometric/spectroscopic observing programs. Only a subset of our brown dwarf searches were specifically tailored to supply targets for our p14034 \textit{Spitzer} campaign. Through the combination of multiple searches described in $\S$\ref{sec:catwise_searches}-\ref{sec:unwise_searches}, we discovered (and visually confirmed) a total of $\sim$2,500 previously unpublished moving objects. Most of these $\sim$2,500 discoveries are not appropriate \textit{Spitzer} targets because they can be sensibly followed up from ground-based facilities. Discussion of additional CatWISE motion discoveries beyond those followed up via \textit{Spitzer} p14034 is deferred to future papers.

We obtained p14034 observations for only the subset of our moving object discoveries which would be extremely difficult to follow up from any platform other than \textit{Spitzer} and/or for which \textit{Spitzer} 3.6$\mu$m (hereafter ch1) and 4.5$\mu$m (hereafter ch2) photometry provides a critically informative diagnostic. We therefore selected only  discoveries falling into one or more of the following categories when constructing our p14034 target list:

\begin{enumerate}
 \item \textbf{Non-stationary objects detected exclusively in W2}. Detectable motion \citep[$\mu \gtrsim 200$ mas/yr at the typical W2 magnitude of our targets;][]{catwise_data_paper} suggests such objects must be relatively nearby. For a moving source, non-detections in W1 and all other shorter wavelengths indicate an extremely cold temperature, low luminosity and hence small distance. Moving objects detected solely in W2 are thus prime close-by Y dwarf candidates, and they were considered the highest priority target class for our \textit{Spitzer} p14034 campaign.
 
 \item \textbf{Moving objects detected in W1 but still potentially within 20 pc}. In some cases the presence of a faint W1 counterpart indicated that a moving object was insufficiently red to plausibly be a Y dwarf. We nevertheless retained such objects for p14034 consideration if the photometric distance estimate implied by W2 and W1$-$W2 indicated that the object could potentially be a new member of the 20 pc sample. Completing the census of objects within this volume is crucial for space density and mass function analyses \citep{davy_parallaxes}. Furthermore, \textit{Spitzer} ch1/ch2 photometry can provide significantly refined photometric distance and spectral type estimates relative to those dependent upon low-significance W1 detections. For these reasons, we deemed $d \lesssim 20$ pc candidates worthy p14034 \textit{Spitzer} targets despite detection in W1.
 
 \item \textbf{Candidate very late type ($\gtrsim$ T0) common proper motion (CPM) companion objects potentially in wide-separation visual binary systems}. Possible CPM status was usually noted serendipitously via our standard visual inspection motion vetting process ($\S$\ref{sec:visualization}), and subsequently checked using the similarity between CatWISE motion of the candidate secondary versus that of the putative primary. \textit{Spitzer} ch1/ch2 observations for CPM candidates can provide a relatively accurate photometric distance  and astrometry for an improved motion estimate. These can then be compared against the distance/motion of the primary to better test the comoving pair hypothesis. Late-type wide CPM companions are rare, so it is also highly valuable to obtain \textit{Spitzer} ch1/ch2 photometric data points for these critical benchmarks.  Presentation of late-type CatWISE CPM discoveries followed up via \textit{Spitzer} p14034 is deferred to a forthcoming paper (Marocco et al., in prep.).
 
 \item \textbf{Objects with exceptionally large proper motion ($\mu \gtrsim 1''$/yr) and/or reduced proper motion\footnote{W2 reduced proper motion is defined as $H_{W2} = m_{W2} + 5 + 5\textrm{log}_{10}(\mu)$, where $\mu$ is the total proper motion in units of arcseconds per year.} ($H_{W2} \gtrsim 19.5$)}. Very high (reduced) proper motion of a source first recognized in the mid-infrared may be indicative of several interesting phenomena, including: a late type subdwarf, a very cold nearby neighbor to the Sun and/or large tangential space velocity.
 
\end{enumerate}

A given object can sometimes fall under more than one of the above target categories. But no moving object discovery was placed on the p14034 program without meeting at least one of the above four selection criteria. All of our targets were, at the time of Astronomical Observation Request (AOR) submission to the p14034 program, never-before-published discoveries; our \textit{Spitzer} campaign was not intended to provide supplementary follow-up of known objects drawn from the literature. Additionally, we vetoed candidates based on checking the \textit{Spitzer} Heritage Archive (SHA) for existing unpublished observations by other teams targeting recent brown dwarf discoveries \citep[e.g., Backyard Worlds;][]{backyard_worlds} so as not to wastefully duplicate \textit{Spitzer} pointings.

In our target selection decision-making, we sought to pursue a ``high-risk, high-reward'' strategy. We declined to place many of our unmistakable late-type moving object discoveries on p14034 because secure W1 detections made clear that they were simply T dwarfs at $\gtrsim 30$ pc. Instead of such ``safe'' objects, we opted to prioritize very faint W2-only candidates with Y dwarf potential, even those so marginal in WISE that they might ultimately turn out to be entirely spurious. The rationale is that mid-T dwarfs don't necessitate \textit{Spitzer}'s unique capabilities in the same way that Y dwarfs do, so we ought to observe as many Y dwarfs as possible before \textit{Spitzer}'s retirement, even at the expense of false positives. Similarly, to find the most superlative targets before \textit{Spitzer}'s retirement, we chose not to limit ourselves to searches based strictly on CatWISE, but instead leveraged additional data products at hand (the AllWISE Catalog and unWISE coadds; see $\S$\ref{sec:allwise_searches} and $\S$\ref{sec:unwise_searches}, respectively).

The following subsections provide a detailed list of our many (14) distinct searches that contributed at least one target to the p14034 photometry presented in Table \ref{tab:wise_spitzer_photom}. For convenience, we refer to each search by a shorthand that consists of one capital letter followed by one digit. Table \ref{tab:wise_spitzer_photom} lists which specific search(es) discovered each target in the ``search method'' column. Our searches generally rely on  motion selection, color selection of objects that are very red in W1$-$W2, or some combination of the two. Each search's candidate identification yielded far more false positives than genuine newly discovered moving objects, so extensive visual inspection campaigns were performed to assess each candidate's possible motion by eye ($\S$\ref{sec:visualization}). The dominant sources of false positives are bright star artifacts, blends and statistical measurement fluctuations that lead to the impression of potentially large motion and/or red W1$-$W2 color.

\subsection{CatWISE-based Selections}
\label{sec:catwise_searches}

Most of our p14034 targets were discovered by directly mining the CatWISE Preliminary Catalog. We ran a total of 11 distinct moving objects searches on this catalog. Note that our searches were performed on early, unfinished versions of CatWISE, rather than the published CatWISE Preliminary Catalog described in \cite{catwise_data_paper}. In many cases, limited or no artifact flagging existed in these early CatWISE databases; this often shaped/constrained our tailoring of search criteria. Also, all CatWISE-based selections were run at a time when there was no distinction between sources now separated into the CatWISE ``catalog'' and ``reject'' tables\footnote{Only two of our p14034 targets are found in the CatWISE Preliminary reject table rather than in the main CatWISE Preliminary catalog: CWISEPR J062436.84$-$071147.2 and CWISEPR J065144.62$-$115106.1.}. Appendix A of \cite{catwise_data_paper} provides column descriptions for the CatWISE quantities involved in our selection criteria.

Our 11 CatWISE-based searches fall under two broad categories: (1) traditional catalog queries each implementing a set of hard cuts ($\S$\ref{sec:catwise_cuts}) and (2) supervised machine learning methods trained on human-verified late-type moving objects ($\S$\ref{sec:catwise_ml}).

\subsubsection{CatWISE Catalog Cuts}
\label{sec:catwise_cuts}

Prior WISE motion surveys such as the AllWISE and AllWISE2 motion searches \citep{allwise_motion_survey, allwise2_motion_survey} were performed via selections cutting on motion, color and artifact flagging catalog columns. Motivated by the success of these previous WISE-based motion surveys, we modeled eight of our search methods after this same general approach, but now applied to the newly available CatWISE data set.

Selection method C1 combines CatWISE color and motion information to isolate objects that are both red in W1$-$W2 and have large W2 reduced proper motion ($H_{W2}$). The usage of $H_{W2}$ rather than simply proper motion itself is a way of prioritizing fast-moving objects of low luminosity, such as Y dwarfs and late type subdwarfs. C1 candidates are obtained from a full-sky CatWISE query requiring (w2snr $ > $ 20), (w2snr\_pm $ > $ 20), ($H_{W2} > 15$), (rchi2/rchi2\_pm $ > $ 1.03), (w2rchi2\_pm $ < $ 2), (W1$-$W2 $ \ge $ 1.5) and $(Q < 10^{-5})$. $Q$ is a significance of motion metric defined in $\S$3.4.1 of \cite{allwise_motion_survey} as $Q = e^{-\chi^2_{motion}/2}$, where $\chi^2_{motion}$ = (pmra/sigpmra)$^2$ + (pmdec/sigpmdec)$^2$. Note that typical WISE sources such as main sequence stars have a color of W1$-$W2 $\approx$ 0 (Vega). The relatively high W2 SNR requirements stipulated as part of this query were necessary to keep the candidate sample size manageable, as this search was performed at a time when no CatWISE artifact flagging was available.

Search C2 is a variant of search C1, but replacing the (W1$-$W2 $ \ge $ 1.5) color cut with a proper motion cut of $\mu > 0.5''$/yr. Again, search C2 was executed without the benefit of any artifact flagging information.

Search C3 implements a combination of the cuts from C1 and C2, and was performed after artifact flagging columns had been added to our working CatWISE database. By leveraging the artifact flagging to remove many spurious candidates, C3 extended to lower W2 SNR than C1/C2, specifically restricting to the range 10 $<$ w2snr\_pm $\le$ 20. C3 also included an additional requirement of (w2snr\_pm $ > $ w1snr\_pm). The significance of motion criterion was made more stringent than in C1 and C2, requiring $Q < 10^{-6}$. Both the proper motion cut $\mu > 0.5''$/yr from C2 and the $H_{W2} > 15$ reduced proper motion cut from C1/C2 were applied. Lastly, search C3 sought to eliminate artifacts by requiring (ab\_flags = `00').

Search C4 is a variant of C1 run after artifact flagging had been put in place. The CatWISE catalog cuts in C4 are the same as those in C1, aside from the following updates: C4 requires (ab\_flags = `00'), flips the rchi2/rchi2\_pm cut to (rchi2/rchi2\_pm $ \le $ 1.03), and adds a new (w1snr\_pm = \verb|null|) W1 non-detection requirement.

Although based on CatWISE catalog cuts, searches C5-C8 are not immediate descendants of the C1-C4 approaches. Search C5 consists of an all-sky CatWISE query requiring $Q < 10^{-5}$, (w2mpro $>$ 13), (w2snr $> $ 10), (k1 = 0) and (k2 = 3). k1 (k2) is a CatWISE column indicating, for each object, which scan direction(s) provided a successful photometric measurement in W1 (W2). k1 = 0 means that neither ascending nor descending WISE scans provided good W1 photometry, while k2 = 3 means that good W2 photometry was obtained for both WISE scan directions. This query is thus, in effect, implementing an alternative means of identifying W2-only moving objects within the CatWISE catalog. Search C5 was also conducted prior to the existence of any CatWISE artifact flagging capabilities.

Search C6 is a full-sky CatWISE query requiring (w2snr $>$ 10), w1flux $<$ (w2flux$\cdot10^{-1.5/2.5} - 2\cdot$w1sigflux), (rchi2/rchi2\_pm $>$ 1.03), (w2rchi2\_pm $<$ 2), $Q < 10^{-6}$, and AllWISE CC flags = `0000' when available. The AllWISE CC flags were gathered via a CatWISE-AllWISE positional cross-match. One notable aspect of this search is that a color cut of effectively (W1$-$W2 $ > $ 1.5) is implemented in terms of fluxes rather than magnitudes, to avoid the complications associated with e.g., quoting magnitudes in cases of zero or negative W1 flux (as can happen for a very red W1 non-detection). In the CatWISE catalog, both w1flux and w2flux have units of Vega nanomaggies\footnote{A source with total flux of 1 Vega nanomaggie has a magnitude of 22.5 in the Vega system.}.

Search C7 is a variant of search C6, but lowering the W2 SNR threshold to (w2snr $>$ 5) in attempt to push fainter. To balance out the large influx of sources at relatively low SNR, the color criterion was made more stringent at effectively (W1$-$W2 $>$ 2.5), with the actual flux-based cut being w1flux $<$ (0.1$\cdot$w2flux$ - 2\cdot$w1sigflux). The significance of motion criterion was loosened to ($Q < 10^{-5}$) for search C7. We additionally required that AllWISE CC flags not contain a capital letter (when an AllWISE cross-match was available) and (ab\_flags = `00').

Search C8 is a pure color selection. We required that candidates not be significantly detected at W1 (w1snr $ < $ 3) and be well-detected in W2 (w2snr~$> 10$) in CatWISE. A negative cross-match (2.5$''$ radius) against AllWISE sources with (w3mpro $<$ 13) was used to remove extragalactic contaminants that tend to be red in W2$-$W3 color. C8 made use of CatWISE ab\_flags to require that candidates not be flagged as W2 ghosts, W2 latents, or W2 diffraction spikes.

\subsubsection{CatWISE Machine Learning}
\label{sec:catwise_ml}
Search method M1 is described fully in $\S$3 of \cite{marocco2019}. In brief, M1 uses the \textit{XGBoost} software package \citep{xgboost} to perform supervised machine learning on the CatWISE catalog. A training set was constructed from the CatWISE sources corresponding to known high proper motion late T and Y dwarfs, with the goal of finding other CatWISE entries displaying similar properties. Search M1 is restricted to CatWISE objects that are faint and red by only classifying the subset of CatWISE rows with (w2mpro $ > $ 14) and also satisfying:

\begin{equation}
\label{equ:dan_frmc}
\begin{multlined}
(w1mpro-(3\times w1sigmpro)) \\ - (w2mpro+(3\times w2sigmpro)) \ge 1.0
\end{multlined}
\end{equation}

This enforces a requirement that each retained source would have a color of (W1$-$W2 $\geq$ 1) even if the CatWISE-reported magnitudes turned out to be 3$\sigma$ bright in W2 and 3$\sigma$ faint in W1.

Search M2 is a variant of search M1, with the classifier trained on a sample including hitherto identified p14034 targets rather than a sample consisting exclusively of previously published late-type brown dwarfs.

Search M3 is also a modified version of search M1, but removing the W1$-$W2 color criterion in Equation \ref{equ:dan_frmc}. The motivation for this variant is the possibility of recovering overlooked late-type moving objects with WISE color measurements corrupted due to blending with background sources, plus the potential to find additional fast-moving sources irrespective of W1$-$W2 color.

\subsection{AllWISE-based Selections}
\label{sec:allwise_searches}

From our prior experiences searching the AllWISE database, we considered it likely that more late T and Y dwarfs remained to be found in that data set, particularly with the aid of our recently upgraded suite of visualization tools used to scrutinize candidates ($\S$\ref{sec:visualization}). Two AllWISE-based moving object searches contributed to the p14034 target list, both utilizing only simple catalog cuts.

In search A1, our candidates are drawn from a full-sky query of the AllWISE catalog, requiring very red W1$-$W2 color (w1mpro$-$w2mpro $ > 3$), relatively little W3 flux (w2mpro$-$w3mpro $ < $ 3.5) to weed out extragalactic contaminants, w2sigmpro not \verb|null|, and W2 CC flags not containing \verb|H|, \verb|O|, \verb|P| or \verb|D| to avoid bright star artifacts. Because of the very extreme W1$-$W2 color cut imposed, this yielded a relatively small sample of candidates, $\sim$500 in total, which were then subjected to visual inspection.

Search A2 likewise identifies candidates using a pure color selection method based on the AllWISE catalog. We retain only those AllWISE rows that are effectively W1 non-detections (w1snr $< 3$), have very red W1$-$W2 colors (w1mpro$-$w2mpro $>$ 2), have ($|b_{gal}| > 10^{\circ}$) to avoid the crowded Galactic plane, (\verb|nb| = 1) to remove blends, and none of \verb|H|, \verb|O|, \verb|P| or \verb|D| in either the W1 or W2 CC flags to discard bright star artifacts. This yielded a sample of $\sim$2,000 candidates spread across the entire $|b_{gal}| > 10^{\circ}$ sky.

\subsection{unWISE-based Selection}
\label{sec:unwise_searches}

One of our selection techniques (search U1) proceeded directly from image-level analysis of the unWISE coadds themselves, rather than the CatWISE or AllWISE catalog. From the time-resolved unWISE coadds \citep{tr_neo3}, we created two full-sky sets of meta-coadds in each band: one built by stacking together all pre-hibernation epochs, and a second built by stacking together all post-reactivation epochs. We subtracted the pre-hibernation W2 meta-coadds from the post-reactivation W2 meta-coadds and ran Source Extractor \citep{source_extractor} on the difference images. The goal was to find sources that moved sufficiently during the $\sim$3 year WISE hibernation period so as to avoid self-subtraction. Source Extractor forced photometry on each of our four sets of meta-coadds (W1 and W2, pre and post hibernation) was performed at the positions of W2 post-reactivation difference image detections. 

The catalog of W2 difference detections, augmented with forced photometry, was then analyzed to select moving object candidates. Specifically, we cross-matched the difference detection catalog with a sample of known late type brown dwarfs to form a training set. An \textit{XGBoost} classifier similar to those described in $\S$\ref{sec:catwise_ml} was then used to identify difference detection catalog entries with properties similar to those of the training sample.

Search U1 was employed because we expected it to perform well for very fast moving sources ($\mu_{tot} \gtrsim 2''$/yr), whereas CatWISE source detection might reasonably fail for extremely faint objects with exceptionally large proper motions. The fact that both of our discoveries presented in this work which lack CatWISE counterparts --- WISEA J153429.75$-$104303.3 ($\mu_{tot} \approx 2.7''$/yr) and WISENF J193656.08+040801.2 ($\mu_{tot} \approx 1.3''$/yr) --- were identified via search U1 attests to the capability of this method to find moving objects that the CatWISE pipeline was not optimized to handle properly.

\subsection{Visualization Tools}
\label{sec:visualization}

Extensive visual inspection of moving object candidates delivered by the searches described in $\S$\ref{sec:catwise_searches}-\ref{sec:unwise_searches} played a vital role in providing a high-purity sample of brown dwarf targets for our \textit{Spitzer} p14034 campaign. In total, we visually inspected $\sim$130,000 candidates in the course of the searches described in $\S$\ref{sec:catwise_searches}-\ref{sec:unwise_searches}. A major factor enabling our discoveries based on CatWISE, AllWISE and unWISE was our usage of visualization tools/aids that leveraged the full W1/W2 time baseline afforded by the combination of pre-hibernation and post-reactivation WISE/NEOWISE imaging.

\subsubsection{Finder Charts}

We created a new, customized version of the multi-panel, multi-wavelength finder chart program used by prior WISE motion searches such as \cite{allwise2_motion_survey} and \cite{allwise2_motion_survey}; for an example see Figure 1 of \cite{schneider_neowise}. We added two sets of W1/W2 time-resolved unWISE coadd cutouts for each candidate, with one set at the beginning of the pre-hibernation WISE mission and one at the end of the third year of the post-reactivation NEOWISE mission. Taking advantage of the $\sim$6.5 year WISE-NEOWISE time baseline in this way allowed us to perceive the source motion (or lack thereof) using only W1/W2 data at widely spaced epochs, rather than needing to obtain an appreciable time baseline by comparison of WISE images to shorter wavelength data sets.

\subsubsection{WiseView Interactive Blinker}
\label{sec:wiseview}

\begin{figure*}
\plotone{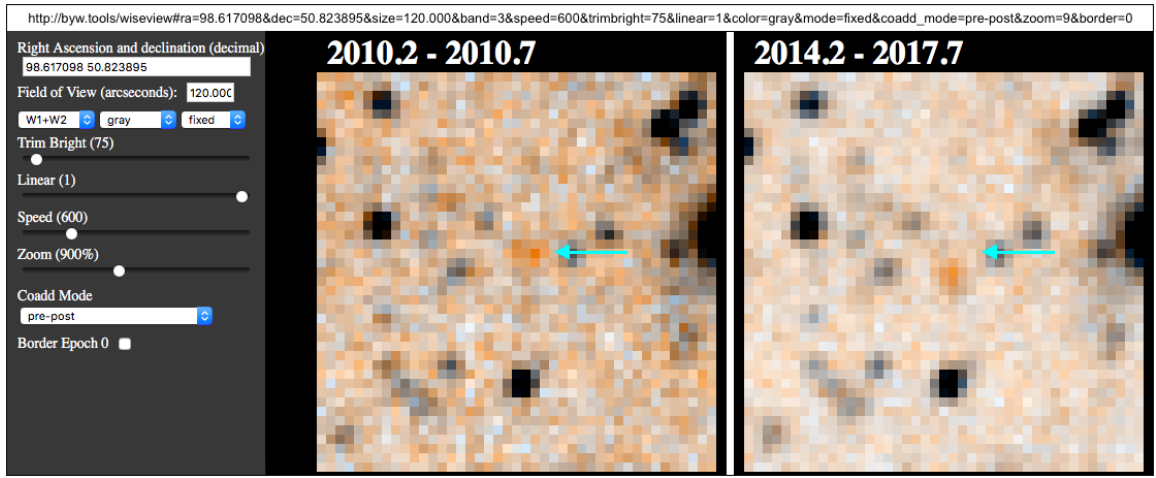}
\caption{Depiction of the WiseView interactive image blinking tool ($\S$\ref{sec:wiseview}) used for visual inspection of potential p14034 targets. At left of the vertical white space is a screen shot of the WiseView interface, including blink tuning widgets and the first of two color composite images created based on the user-specified blink parameters. This blink sequence is centered on CWISEP J063428.10+504925.9, the orange source indicated by the cyan arrow. The cyan arrow remains at the same sky position in both panels; it is not present in WiseView, but rather has been edited in to highlight the large southeasterly motion of this object. The ``pre-post'' Coadd Mode selected generates two meta-coadds per band, with the first spanning the full pre-hibernation time period (2010.2-2010.7 in this case) and the second spanning the available post-reactivation time period (2014.2-2017.7 in this case). A screenshot of the post-reactivation WiseView color composite is included at right of the vertical white space, with the WiseView widget panel omitted. Although these two images are shown side-by-side here, WiseView presents a single animated blink alternating between these two images, with both images aligned in the same position on the screen. The parameters of this blink are encoded in a WiseView URL provided at top --- one can experience this WiseView animation in action by visiting that URL.}

\label{fig:wiseview_example}
\end{figure*}

Our visual inspection workflow relied heavily on a new visualization tool called WiseView \citep{wiseview} not previously available for WISE moving object searches such as the AllWISE and AllWISE2 motion surveys. In contrast to static multi-wavelength finder charts, WiseView is an interactive browser-based interface for creating customized animated blinks of time-resolved unWISE coadds. Numerous blink parameters are tunable in real time, including the band(s) shown (W1, W2, or both), the central sky position, the frame rate, the stretch and the field of view size. Figure \ref{fig:wiseview_example} illustrates an example of the WiseView interface as employed when vetting one of our W2-only brown dwarf discoveries.

\subsubsection{DESI Imaging Viewer}

To select moving objects detected in W2 but not at any shorter wavelengths, we made extensive use of the DESI pre-imaging ``Legacy Surveys'' sky viewer\footnote{\url{http://legacysurvey.org/viewer}} to inspect red-optical survey images. This viewer allows for interactive exploration of wide-area survey data sets with deep $z$ and $Y$ band imaging, in particular DECaLS/MzLS \citep[$z \approx 23.0$ AB at $5\sigma$ over $\sim$1/3 of the sky;][]{dey_overview} and Dark Energy Survey DR1 \citep[$z \approx 23.4$ AB and $Y \approx 22.2$ AB at $5\sigma$ over $\sim$1/8 of the sky;][]{des_dr1}. Visible $z$ and/or $Y$ counterparts were generally treated as evidence that a moving object candidate was either insufficiently red to be a Y dwarf or else extragalactic if the red-optical counterpart appeared extended.

\subsubsection{IRSA Finder Chart}

In some cases where conclusively confirming/denying motion by eye proved difficult, we consulted AllWISE W3 and W4 images via the IRSA Finder Chart application\footnote{\url{https://irsa.ipac.caltech.edu/applications/finderchart/}}. Because our motion candidates are so faint (median W2 $\sim$ 15.9; see $\S$\ref{sec:sample_properties}), a strong counterpart at W3 and/or W4 would only be expected in the case of a stationary extragalactic source but not for a late-type brown dwarf. We therefore avoided selecting sources seen to have coincident W3/W4 emission as p14034 targets.

\subsubsection{PanSTARRS-1 Cutouts}
\label{sec:ps1_cutouts}

Before placing candidates on the p14034 target list, we inspected PanSTARRS-1 image cutouts\footnote{\url{http://ps1images.stsci.edu/cgi-bin/ps1cutouts}} for objects north of $\delta \approx -30^{\circ}$ \citep{the_ps1_surveys}. Visible detections in PanSTARRS-1 were in general used to veto potential Y dwarf candidates --- given the faintness of our sample in W2, detection in any PanSTARRS-1 filter would be inconsistent with a Y dwarf spectral type.

\section{General p14034 Sample Properties}
\label{sec:sample_properties}

\begin{figure}
\plotone{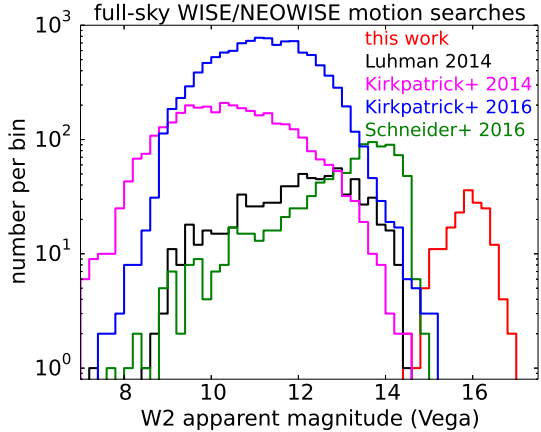}
\caption{Comparison of W2 magnitude distributions for moving object discoveries from published full-sky WISE/NEOWISE motion surveys. Each histogram shows the number of new discoveries per 0.2 mag W2 bin for a given sample. Note the logarithmic scale. Red is our entire 174-object \textit{Spitzer} p14034 target list (including 4 CPM targets and 6 spurious candidates from $\S$\ref{sec:duds}) plus our 3 brown dwarf candidates with ch1/ch2 photometry based on archival \textit{Spitzer} imaging. Our discoveries are much fainter than those of previous full-sky WISE/NEOWISE motion searches.}
\label{fig:w2_histogram}
\end{figure}

We filled our allocated 40.5 hours of Cycle 14 \textit{Spitzer} time with 174 unique brown dwarf candidate targets, each of which received a single corresponding AOR as detailed in $\S$\ref{sec:observing_strategy}. As discussed in $\S$\ref{sec:target_selection}, these 174 targets represent only a small subset of the visually vetted moving object discoveries yielded by our searches. Figure \ref{fig:w2_histogram} shows that our targets are much fainter than those of previous WISE motion surveys, including the AllWISE and AllWISE2 searches \citep{allwise_motion_survey, allwise2_motion_survey}. The median W2 magnitude of our targets is 15.93, with a dispersion of 0.47 mags. For comparison, the W2 single-exposure depth is W2 $\approx$ 14.5, which has represented the faint limit of prior WISE-based motion searches \citep[e.g.,][]{luhman_planetx, schneider_neowise}. Figure \ref{fig:w2_snr_histogram} shows the W2 SNR distribution of our \textit{Spitzer} photometry sample. The median (mean) W2 SNR is just 16.2 (18.4), with a dispersion of 8.6. Note that these values indicate the total W2 SNR when combining 4 years of WISE/NEOWISE imaging; detections in any time slice of the available W2 imaging will generally be of even lower significance.

Figure \ref{fig:spatial_distribution} shows the spatial distribution of our p14034 targets, which, as expected, are scattered fairly uniformly across the entire sky while preferentially avoiding the confused Galactic plane.

\subsection{Candidates with Archival ch1/ch2 Data}
Three of our discoveries not previously recognized as brown dwarf candidates happened to have sufficient serendipitous archival ch1/ch2 imaging in SHA to enable robust phototyping without the need for additional p14034 observations. We performed our usual \textit{Spitzer} photometry ($\S$\ref{sec:spitzer_photometry}) and astrometry ($\S$\ref{sec:spitzer_astrometry}) on these archival observations to obtain ch1$-$ch2 colors. These three objects (CWISEP 0229+7246, CWISEP 1721+5950 and CWISEP 2247$-$0041) are denoted by blue squares in Figure \ref{fig:spatial_distribution}.

\subsection{Targets as yet Unobserved by \textit{Spitzer}}

Three of our p14034 brown dwarf targets are scheduled for \textit{Spitzer} photometry in the near future but remain unobserved by \textit{Spitzer} as of this writing (2019 October; CWISEP 1353$-$0037, CWISEP 1515$-$2157 and CWISEP 0601$-$5922). In all three such cases, the motion is conclusively confirmed with WISE astrometry alone, so we have chosen to present these three discoveries in this paper despite the current lack of available \textit{Spitzer} photometry/astrometry. These three targets are therefore listed in Table \ref{tab:wise_spitzer_photom}, but with \textit{Spitzer} photometry columns left empty.

\begin{figure}
\plotone{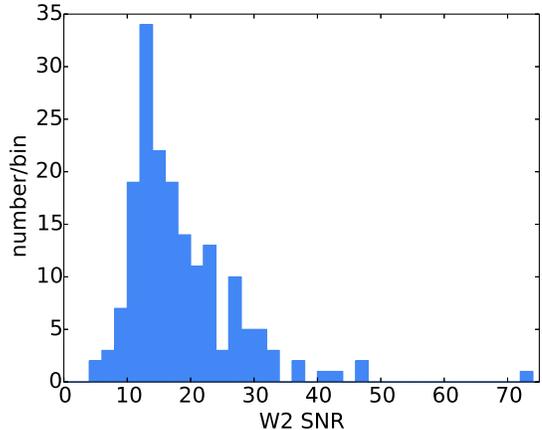}
\caption{W2 signal-to-noise distribution for our moving object discoveries characterized with \textit{Spitzer}. The histogram shows our entire 174-object \textit{Spitzer} p14034 target list (including 4 CPM and 6 spurious candidates from $\S$\ref{sec:duds}) plus our 3 brown dwarf candidates with ch1/ch2 photometry based on archival \textit{Spitzer} imaging. The median (mean) W2 SNR is 16.2 (18.4), with a dispersion of 8.6.}
\label{fig:w2_snr_histogram}
\end{figure}

\begin{figure}
\plotone{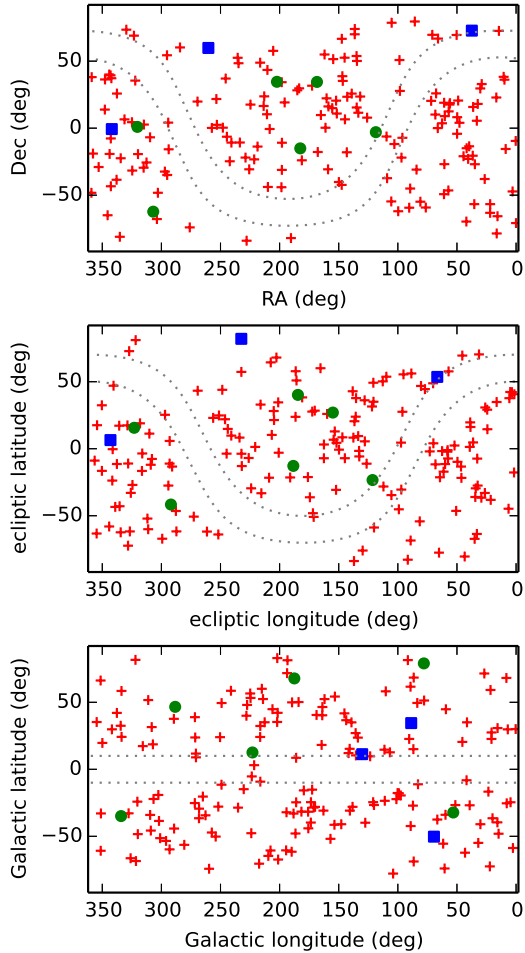}
\caption{Spatial distribution of late-type brown dwarf candidates on p14034 or for which archival \textit{Spitzer} data yielded a ch1$-$ch2 color. CPM candidates have been removed, since those are presented separately in a forthcoming paper. Red plus marks are p14034 targets. Green circles are spurious p14034 candidates (see Table \ref{tab:duds}). Blue squares are brown dwarf candidates for which our ch1$-$ch2 colors are based on serendipitously available archival \textit{Spitzer} imaging. Top: equatorial coordinates. Middle: ecliptic coordinates. Bottom: Galactic coordinates. In all cases the dotted grey lines denote Galactic latitude of $\pm$10$^{\circ}$}
\label{fig:spatial_distribution}
\end{figure}

\section{\textit{Spitzer} Observing Strategy}
\label{sec:observing_strategy}

We based our observing strategy on those of prior \textit{Spitzer} campaigns designed to measure the colors of WISE-selected late T and Y dwarf candidates (e.g., programs 70062 and 80109; PI Kirkpatrick). These foregoing \textit{Spitzer} programs generally observed substantially brighter objects than those comprising our p14034 target list, and showed that coaddition of five 30 second ch1 dithers typically achieves signal-to-noise of 10 (5) at a Vega magnitude of ch1 = 18.0 (18.75).

The boundary between the ch1$-$ch2 colors of late T and Y dwarfs occurs at ch1$-$ch2 $\approx$ 2.4 \citep[e.g.,][]{davy_parallaxes}, and our primary goal was to obtain ch1 imaging deep enough to distinguish between late T and Y dwarfs. We therefore attempted to engineer our number of ch1 dithers such that a $5\sigma$ ch1 detection would always establish a color of at least ch1$-$ch2 = 2.75 magnitudes.

For each target, we chose either 5, 7, 9, 11, or 13 ch1 dithers (at 30 seconds per dither) so that we would achieve ch1 SNR of at least 5 for ch1$-$ch2 = 2.75, under the assumption that the ch2 mag would be equal to the W2 mag. According to this strategy, W2 $\le$ 16 candidates receive five ch1 dithers, and the break point between 5 and 7 dithers is W2 = 16.0. Then the break point between 7 and 9 dithers occurs at W2 = 16.18, and so forth. We exercised some case-by-case discretion in bumping up the exact number of dithers chosen, but always used ch1 SNR of 5 at ch1$-$ch2 = 2.75 to enforce a minimum number of ch1 dithers.

For each target, both ch1 and ch2 were obtained as part of the same AOR. This minimizes slew overheads and ensures that our color measurements cannot be corrupted by any long timescale variability aliasing into images of the same target taken at large time separation. We always acquired the same number of 30 second dithers in both ch1 and ch2 for each target. This effectively maintains a fixed ch2 SNR of $\sim$75 for the high-significance ch2 detection across all members of our sample, which ensures the high value of each target's ch2 detection for astrometry ($\S$\ref{sec:spitzer_astrometry}). We dithered with a `cycling' pattern of medium scale\footnote{\url{https://irsa.ipac.caltech.edu/data/SPITZER/docs/irac/calibrationfiles/dither/}}.

The \textit{Spitzer} p14034 imaging analyzed in this work was acquired between 2018 October 21 and 2019 August 16.

\section{\textit{Spitzer} Photometry}
\label{sec:spitzer_photometry}

All of our \textit{Spitzer} photometry/astrometry analyses presented in this work are based on custom mosaics built from the single-dither BCD images with MOPEX \citep{mopex, mopex_extraction}. We created one such custom mosaic per band (ch1, ch2) per AOR. Relative to using the default ``PBCD'' mosaics supplied for each AOR via SHA, creating our own custom mosaics provided us extra freedom, for example to reject occasional problematic single-frame IRAC images with e.g., a cosmic ray contaminating the targeted brown dwarf candidate. The algorithmic rejection of single-frame outliers such as cosmic rays also appears to be much better overall in our custom mosaics than in the PBCD stacks. In a handful of cases, we excised one or two BCD frames from our custom mosaics due to the presence of a cosmic ray contaminating the faint ch1 counterpart (CWISEP 0035$-$1532, CWISEP 0403$-$4916, CWISEP 1434$-$1344, CWISEP 2251$-$0740 and CWISEP 2355+3804).

Extraction and photometry of sources within our custom \textit{Spitzer} mosaics proceeded as described in $\S$4 of \cite{marocco2019}. In brief we use the MOPEX/APEX software to detect sources in our custom mosaics and perform both PRF-fit and aperture photometry (including the application of an aperture correction) in each of ch1 and ch2. Note that our photometry was always run independently in ch1 and ch2 --- we did not employ a forced photometry approach. By default we used an SNR = 5 source detection threshold. In the case of two very red objects (CWISEP 1434$-$1344 and CWISEP 1446$-$2317), obtaining a ch1 counterpart extraction required lowering the ch1 detection threshold to SNR = 2.

Table \ref{tab:wise_spitzer_photom} lists our ch1 and ch2 photometry results. The quoted magnitude values and their uncertainties are derived by averaging the aperture-based and PRF-fit quantities for each target in each \textit{Spitzer} band. Table \ref{tab:wise_spitzer_photom} also reports the W1 and W2 photometry for each target. With only a small number of exceptions, such as our two discoveries not detected by CatWISE, this WISE magnitude information is drawn from the CatWISE columns w1mpro\_pm, w1sigmpro\_pm, w2mpro\_pm, and w2sigmpro\_pm.

\subsection{Spurious Candidates}
\label{sec:duds}
A small number of our p14034 targets turned out to be entirely absent in the deeper, higher-resolution \textit{Spitzer} imaging we obtained (i.e., a \textit{Spitzer} counterpart could neither be extracted nor visually identified). Table \ref{tab:duds} lists such cases (6/174 = 3.4\% of our targets). Throughout this paper, these 6 completely spurious targets are omitted from various tabulations and analyses, particularly those such as the ch1/ch2 photometry listed in Table \ref{tab:wise_spitzer_photom} that would require a \textit{Spitzer} detection.

We note that \cite{catwise_data_paper} quotes a CatWISE reliability of just under 98\% at W2 = 16. Given that our sample's median magnitude is W2 = 15.93, a 3.4\% rate of spurious sources is within reason.

\subsection{Candidates with Two \textit{Spitzer} Counterparts}
\label{sec:spitzer_pairs}

In two cases, our WISE-based brown dwarf candidate turned out to have two distinct, closely spaced \textit{Spitzer} ch2 counterparts. CWISEP 1541+5230 has two \textit{Spitzer} counterparts in both ch1 and ch2, whereas CWISEP 0229+7246 has two \textit{Spitzer} counterparts in ch2 but only one blended/elongated \textit{Spitzer} counterpart in ch1. In these cases we label the two components by adding a suffix of either ``N'' (northern) or ``S'' (southern) to their designations, based on their relative \textit{Spitzer} ch2 (RA, Dec) positions. In both of these cases, it remains plausible that there are simply two static \textit{Spitzer} counterparts corresponding to our single WISE target, so we vetoed these candidates from being formally considered motion-confirmed in downstream analyses.

\section{Astrometry}
\label{sec:astrometry}

We seek to use motion as a proxy for confirming an object is a nearby brown dwarf. High-significance motion establishes solar neighborhood membership, whereas objects consistent with remaining stationary may be of e.g., extragalactic origin. Because we lack spectroscopic confirmations and our most interesting p14034 targets are detected only in WISE and \textit{Spitzer}, detailed astrometric analysis is needed to best determine whether each source is indeed moving. In this section we explain how we have combined astrometry from both WISE and \textit{Spitzer} to best identify the subset of our brown dwarf candidates that have statistically significant proper motions. The inclusion of \textit{Spitzer} astrometry is a critical component of this analysis, since our \textit{Spitzer} data point provides a completely independent cross-check on the perceived WISE-based motion used to select our candidates; if the \textit{Spitzer} detection ``lines up'' along the WISE astrometric trajectory, then this gives us strong reassurance that the candidate was selected due to true motion rather than a rare fluke in the WISE data.

Using the methodology described here, we limit our astrometric analysis to fits of apparent linear motion; we do not seek to obtain/constrain the parallaxes of our targets. In general, fitting parallaxes for brown dwarfs as faint as our targets will require multiple epochs of \textit{Spitzer} (or similarly precise) observations sampling both sides of the parallactic ellipse \citep[e.g.,][]{davy_parallaxes}, whereas our p14034 imaging provides just a single \textit{Spitzer} astrometric epoch. Additionally, as discussed in $\S$\ref{sec:wise_astrometry}, we typically must coadd data acquired from both sides of WISE's orbit in order to obtain W2 detections of our exceedingly faint targets, meaning that our WISE astrometry is generally unsuitable for parallax fitting.

Our astrometric analysis incorporates ch2 and W2, but never ch1 or W1. This is because our targets are very red in both ch1$-$ch2 and W1$-$W2. For our p14034 \textit{Spitzer} imaging, where both bands received the same total exposure time, the SNR of each target's ch2 detection will be much higher than that of its ch1 detection. Further, because the ch1 and ch2 images of a given target are nearly contemporaneous, folding in ch1 does not offer the possibility of an appreciably extended time baseline, and in combination with ch2 astrometry would merely lead to a negligible improvement in the p14034 \textit{Spitzer} positional precision. The same considerations apply with regard to W1: data acquisition in W1 and W2 is simultaneous, so W1 astrometry would provide only a set of much less precise positions at the same epochs as our W2 astrometric data points.

All \textit{Gaia}-recalibrated W2 and ch2 (RA, Dec) coordinates quoted throughout this paper are in ICRS. The \textit{Spitzer} and WISE positions reported in Tables \ref{tab:wise_positions} and \ref{tab:spitzer_positions} are relative rather than absolute --- we did not attempt to correct for the typically very small parallaxes of our astrometric calibration sources.

\subsection{Gauging Significance of Motion}
\label{sec:chi2_motion}

There are various ways one could imagine quantifying significance of motion. In this work, we opt for a simple, intuitive metric that has been applied in the course of past WISE-based moving object analyses:

\begin{equation}
\chi^2_{motion} = (\mu_{\alpha}/\sigma_{\mu_{\alpha}})^2 + (\mu_{\delta}/\sigma_{\mu_{\delta}})^2
\end{equation}

This $\chi^2_{motion}$ statistic has previously been used during e.g., the AllWISE and AllWISE2 motion surveys \citep{allwise_motion_survey, allwise2_motion_survey}. $\chi^2_{motion}$ tends to increase with larger (absolute) linear motion components and also with decreasing uncertainties on the linear motion measurements. It can also be thought of as corresponding to a false alarm rate, $Q = e^{-\chi^2_{motion}/2}$, where $Q$ is the probability of a statistical fluke causing $\chi^2_{motion}$ to be exceeded.

In this work, we set the threshold for WISE+\textit{Spitzer} `motion confirmation' at $Q = 10^{-5}$, which corresponds to $\chi^2_{motion}$ = 23.03. Ignoring the relatively high-precision \textit{Spitzer} astrometric data points available from p14034 follow-up, a substantial fraction of our sample's targets (46\% = 79/173) have CatWISE $\chi^2_{motion}$ less than this threshold, illustrating the critical need to combine ch2 and W2 astrometry toward better confirming/refuting source motions.

\subsection{Strategy for Combining WISE \& \textit{Spitzer} Astrometry}

Given that CatWISE linear motion estimates are almost always available for our targets, one might imagine concocting a scheme to combine these with our \textit{Spitzer} astrometric data points and thereby derive high-quality WISE+\textit{Spitzer} linear motions. There are many reasons why we find this approach undesirable, for instance:

\begin{itemize}
    \item It isn't entirely clear how to properly combine a CatWISE motion estimate and a \textit{Spitzer} position in order to obtain a $\chi^2_{motion}$ value. 
    \item Two of our most exciting discoveries are absent from the CatWISE catalog (WISEA 1534$-$1043, WISENF 1936+0408). Another two targets have `null' motion uncertainties in CatWISE (CWISEP 0402$-$2651, CWISEP 0430+2556). So in any event, we need to develop an alternative motion-fitting methodology not reliant on CatWISE to address this subset of our targets.
    \item In cases when CatWISE linear motions are corrupted by blending at some subset of WISE epochs, this can be circumvented by careful subselection of epochal W2 detections.
    \item CatWISE used an ad hoc scaling of unWISE pixel-level uncertainties, which could lead to non-optimal CatWISE motion uncertainty estimates.
    \item CatWISE only incorporated NEOWISE data through 2016, whereas additional NEOWISE data are now available.
    \item CatWISE fits W1 and W2 simultaneously --- any nonzero weighting of W1 data in CatWISE motion fits will essentially have added noise to its motion measurements for our W2-only sources.
\end{itemize}

The alternative approach we prefer is to extract our own epochal WISE source catalogs and use these to assemble a vetted list of high-quality W2 astrometric detections for each brown dwarf candidate. For each target, its set of W2 detections can then be straightforwardly combined with our \textit{Spitzer} astrometric data point via simple least squares fitting of apparent linear motions in each of RA and Dec. Moreover, the carefully assembled lists of W2 positions derived during this process may be of substantial interest in their own right, as they can be  combined with any future astrometric follow-up acquired. $\S$\ref{sec:wise_astrometry} ($\S$\ref{sec:spitzer_astrometry}) explains in detail how we obtain the WISE W2 (\textit{Spitzer} ch2) astrometric detections for our targets.

\subsection{WISE Astrometry}
\label{sec:wise_astrometry}

It is challenging to obtain a time series of WISE astrometric detections for moving sources as faint as our targets. By selection, our brown dwarf candidates tend to be completely undetected in W1. In W2, they typically have SNR of just $\sim$15 even when combining four years of WISE/NEOWISE data (see Figure \ref{fig:w2_snr_histogram}). As a result, it is almost never possible to extract single-exposure W2 astrometry for any target in our sample, and we do not attempt to do so. Furthermore, in most cases it is not possible to obtain W2 detections of our targets even in  time-resolved unWISE coadds that stack together the $\gtrsim$12 W2 exposures at each sky location during each single six-monthly WISE sky pass. Therefore, we often must perform source extraction on W2 stacks that combine multiple WISE sky passes.

The latest full-sky unWISE data release \citep{neo4_coadds} provides time-resolved coadds that bin W2 exposures into a series of single WISE sky passes, incorporating both the pre-hibernation time period (2010-2011) and the first four years of NEOWISE-R observations (2013-2017). These time-resolved unWISE coadds form the starting point for our W2 astrometry analysis. Ideally, we would also have access to such unWISE coadds for the calendar 2018 time period, but these have not yet been generated. However, since the 2018 NEOWISE-R data is only slightly earlier in time than our \textit{Spitzer} p14034 imaging, there would be relatively little marginal benefit attained by including 2018 W2 data --- this additional W2 imaging would not increase our overall WISE+\textit{Spitzer} time baseline and would only contribute a relatively weak astrometric constraint adjacent in time to our much higher precision p14034 \textit{Spitzer} data point. With the \cite{neo4_coadds} unWISE data set, we typically have five years of W2 imaging available at each sky location, corresponding to 10 time-resolved W2 coadds, which are labeled with names e000, e001, ..., e009. For concreteness, one representative cadence of such time-resolved unWISE coadds at fixed sky location is e000 $\sim$ 2010.4, e001 $\sim$ 2010.9, e002 $\sim$ 2014.4, ..., e009 $\sim$ 2017.9.

In gathering astrometric detections for our brown dwarf candidates, we always attempt extractions from all  available e??? W2 unWISE coadds. Because this still leaves many of our faint targets undetected, we also generate and perform source extraction on a set of W2 `meta-coadd' time slices. This set of meta-coadd `slices' is listed in Table \ref{tab:slices}. The `pre' slice stacks all pre-hibernation e??? W2 unWISE coadds together to form a deeper 2010-2011 meta-coadd. Analogously, the `post' slice stacks all post-hibernation e??? W2 unWISE coadds together, resulting in a deep 2013-2017 coadd. The `post?$\_$1yr' meta-coadd slices stack the post-hibernation e??? W2 unWISE coadds within a series of four non-overlapping 1-year time intervals. Lastly, the `post?$\_$2yr' meta-coadd slices stack the post-reactivation e??? W2 unWISE coadds within a series of two non-overlapping 2-year time intervals.

We generate our W2 meta-coadds for the full $1.56^{\circ} \times 1.56^{\circ}$ unWISE coadd tile footprint containing each target. This has multiple advantages relative to considering only small postage stamps about our targets. First, it allows us to obtain a large number of \textit{Gaia} DR2 \citep{gaia_dr2} calibrator sources with which to produce refined WCS solutions for each W2 time slice ($\S$\ref{sec:w2_recalibration}). Second, it provides sufficient numbers of bright point sources to accurately model the W2 PSF within each time slice.

We create W2 meta-coadds by performing an inverse variance weighted sum of the contributing e??? unWISE coadds, making use of the unWISE \verb|-invvar-m| inverse variance maps. Using these same inverse variance weights, we also create a corresponding map of the mean MJD for each meta-coadd, enabling us to quote MJD values corresponding to our W2 astrometric detections.

Our modeling of the time-resolved unWISE coadds and  meta-coadds, including source detection and centroiding, was performed using the \verb|crowdsource| crowded field photometry pipeline \citep{unwise_catalog}. \verb|crowdsource| has proven adept at modeling unWISE W1 and W2 images during creation of the full-sky unWISE Catalog \citep{unwise_catalog}. \verb|crowdsource| derives a PSF model for each unWISE image it processes, and reports profile-fit astrometry that is equivalent to flux-weighted centroiding because the nominal PSF center is defined to coincide with the PSF model's flux-weighted centroid.

For the unWISE tile footprint containing each target, we ran \verb|crowdsource| on all W2 time-resolved coadds (e??? time slices) and all meta-coadds (Table \ref{tab:slices} time slices). Next, we proceeded to select the subset of these \verb|crowdsource| detections that are counterparts to each brown dwarf candidate and will ultimately be combined with \textit{Spitzer} astrometry during our final linear motion fits. We began by identifying a visually vetted set of pre-hibernation W2 \verb|crowdsource| counterparts, one for each target. In combination with our \textit{Spitzer} positions ($\S$\ref{sec:spitzer_astrometry}), this allowed us to bracket each target's $\sim$2010-2019 trajectory and derive a crude linear motion estimate. Using this preliminary motion estimate, we then identified all \verb|crowdsource| detections near the moving object's trajectory in all time slices. We visually inspected all such potential counterparts, removing severe blends and static contaminants.

The last step in selecting \verb|crowdsource| detections for our final WISE+\textit{Spitzer} motion fits is to pare down the full set of available detections for each target into a list that incorporates information from each WISE sky pass exactly once\footnote{In some rare cases, such as moving objects severely blended with static contaminants at certain WISE epochs, it was not possible (or desirable) to achieve this idealized goal.}. In this context, our faintest targets are the simplest. These objects will only have W2 detections in the deepest meta-coadds on each side of the WISE hibernation boundary: the `pre' and `post' time slices. Indeed, the simple combination of `pre' and `post' \verb|crowdsource| astrometry was adopted for 104 of our 167 targets\footnote{In the course of this work, we consider a total of 177 moving object candidates: 174 from our p14034 campaign and 3 with archival \textit{Spitzer} imaging. Table \ref{tab:wise_positions} omits 4 of these discoveries that will be presented in an upcoming paper on CatWISE CPM systems (Marocco et al., in prep.) and another 6 targets which turned out to be spurious ($\S$\ref{sec:duds}), leaving 167 objects in Table \ref{tab:wise_positions}.}, as can be seen in Table \ref{tab:wise_positions}. For brighter targets, we have the freedom to choose the specific set of time slices adopted. For instance, bright targets will have e000, e001 and `pre' time slice detections available. We cannot use all of these in our joint WISE+\textit{Spitzer} fits, since this would effectively double-count W2 imaging during the 2010-2011 time period. In these situations, we carefully constructed lists of \verb|crowdsource| detections for each target that omitted as few W2 sky passes as possible while never double-counting. In doing so, we enforced a preference for shorter time slices, with the rationale being that longer time slices incur more smearing of the moving object, which is non-optimal for centroid measurements. Table \ref{tab:wise_positions} lists the W2 time slices employed for each target in our production WISE+\textit{Spitzer} linear motion fits. On average, $\sim$3 W2 detections per target are used.

\subsubsection{W2 Astrometric Recalibration}
\label{sec:w2_recalibration}

The time-resolved unWISE coadds inherit low-level (up to a few hundred mas) astrometric systematics by virtue of propagating the single-exposure WISE astrometry without modification \citep{tr_neo2, neo4_coadds}. We therefore sought to improve the accuracy of our \verb|crowdsource| W2 astrometry by recalibrating it to \textit{Gaia} DR2.

For the \verb|crowdsource| catalog corresponding to each time slice of each unWISE tile footprint, we seek to compute a scalar astrometric offset along each sky direction so as to bring our W2 centroids into best agreement with \textit{Gaia} DR2. In practice, this is accomplished by adding a small offset to each of the two CRPIX components in the native W2 unWISE coadd WCS.

Each coadd from which we draw a brown dwarf candidate W2 detection covers a $\sim$2.4 square degree sky area, resulting in an abundance of available \textit{Gaia} DR2 calibrators. To assemble a set of \textit{Gaia}-\verb|crowdsource| calibration sources, we crossmatch the full list of \verb|crowdsource| detections against a subset of \textit{Gaia} DR2 rows using a 2$''$ radius. We require that our \textit{Gaia} DR2 astrometric calibrators have \textit{Gaia} proper motions available, and we use these to propagate each calibrator's position to the mean epoch of the W2 coadd under consideration. We do not attempt to correct for the \textit{Gaia} calibrator parallaxes, which have a mean amplitude of only $\sim$1-2 mas. The median number of \textit{Gaia} DR2 calibrators employed per W2 coadd is 6,880.

With \textit{Gaia} positions at the relevant epoch in hand, we calculate the two-element shift that needs to be applied to the native CRPIX to zero out the median offsets between the \textit{Gaia} calibrators and their \verb|crowdsource| matches along the coadd $x$ and $y$ pixel directions. We then apply this offset to create a slightly modified CRPIX value that, in combination with the other native W2 WCS parameters, provides a recalibrated astrometric solution most consistent with \textit{Gaia} at zeroth order.

The mean amplitude of the per-coordinate offsets applied to the native CRPIX values is $\sim$50 mas. Using the recalibrated WCS for each coadd, the typical bright end scatter (assessed with  $10 < $ W2 $ < 11.6$ unsaturated sources) relative to \textit{Gaia} `truth' is just 44 mas (41 mas) in RA (Dec). This is a very small fraction of the $\sim$6.5$''$ W2 PSF FWHM ($\sim$1/150 FWHM), providing confidence in the astrometric fidelity of our W2 (meta-)coadds.

\begin{figure*}
\plotone{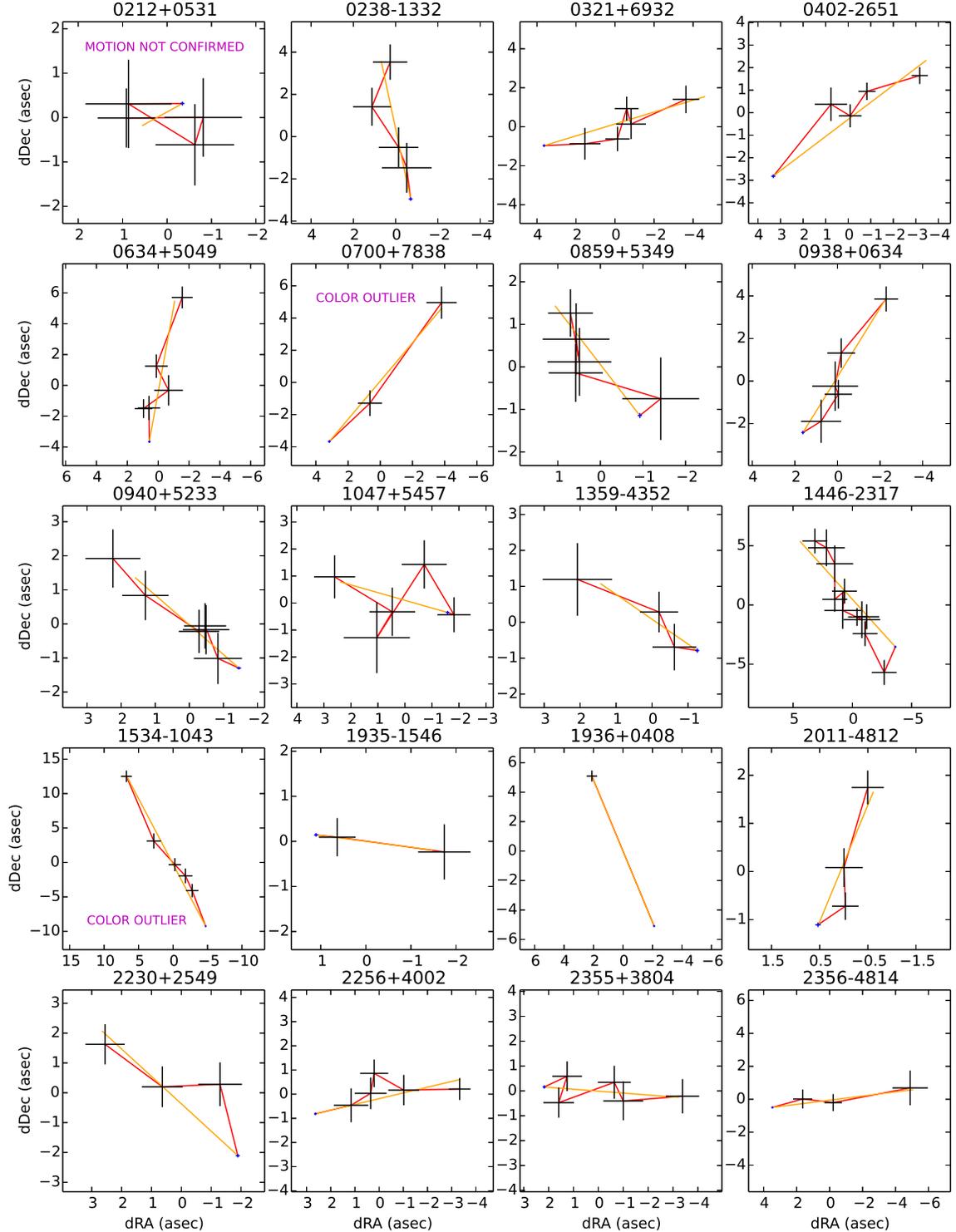}
\caption{Combined WISE+\textit{Spitzer} astrometric trajectories for all 18 of our discoveries with best-fit ch1$-$ch2 color most consistent with type $\ge$Y0. Except for CWISEP 0212+0531, all of these brown dwarf candidates are motion-confirmed according to our $\chi^2_{motion}$ criterion. Also included are two fast-moving ($\mu_{tot} > 1''$/yr) targets with unusually large $J - $ch2 colors relative to their ch1$-$ch2 colors: WISEA 1534$-$1043 and CWISEP 0700+7838 (see Figure \ref{fig:j_minus_ch2}). The black plus marks are centered on the WISE detections, with line segments extending $\pm$1$\sigma$. The much smaller blue plus marks provide the same information for each target's \textit{Spitzer} p14034 astrometric data point. Red lines connect positions adjacent to one another in time. Orange lines show the best-fit linear motion trajectory (Table \ref{tab:wise_spitzer_pm}) over the same time period spanned by the combination of WISE and \textit{Spitzer} data points.}
\label{fig:radec_plots}
\end{figure*}

The systematics floor of our W2 astrometry as characterized by the bright end scatter is very small compared to the typical per-coordinate statistical uncertainties on our W2 \verb|crowdsource| centroids, which have a median value of 515 mas. The W2 centroid statistical uncertainties are large because of the broad W2 PSF and  low SNR of our \verb|crowdsource| W2 detections (median SNR = 9.8). Figure \ref{fig:radec_plots} provides a visual illustration of the large statistical noise inevitably present in our W2 centroids.

Table \ref{tab:wise_positions} provides the positions and associated metadata of all W2 detections that feed into our final WISE+\textit{Spitzer} linear motion fits. The $\sigma_{RA}$ and $\sigma_{Dec}$ values quoted each result from summing the statistical centroid uncertainty and bright end scatter systematics floor in quadrature (even though the former strongly dominates over the latter). $\sigma_{RA}$ is in angular rather than coordinate units.

\subsection{\textit{Spitzer} Astrometry}
\label{sec:spitzer_astrometry}

Whereas our typical per-coordinate W2 centroid precision is larger than 500 mas, we should expect the ch2 per-coordinate centroid precision to be characteristically an order of magnitude smaller. This is simply due to the much higher SNR of our target detections in ch2 (median SNR $\approx$ 77) than in W2 (median SNR $\approx$ 10) and $\sim$3$\times$ narrower ch2 PSF. Because our centroid uncertainties are so much smaller in ch2 than in W2, the uncertainties on our eventual WISE+\textit{Spitzer} linear motion measurements will be almost entirely dictated by the W2 positional uncertainties. Thus, we need not become fixated on obtaining \textit{Spitzer} astrometry that achieves the ch2 imaging's theoretically optimal precision floor.

\cite{davy_parallaxes} described a procedure for measuring \textit{Spitzer} ch2 astrometry while achieving a systematics floor of just 10 mas per coordinate. This entailed a rather involved analysis which proceeds directly from the set of individual \textit{Spitzer} BCD frames contributing to each AOR. On the other hand, \cite{martin_spitzer_astrometry} demonstrated that a systematics floor of $\sim$20 mas per coordinate could be achieved via a substantially more convenient procedure based on astrometric measurements performed on MOPEX mosaics. Because reducing our ch2 systematics floor from $\sim$20 mas to $\sim$10 mas would negligibly decrease the uncertainties on our WISE+\textit{Spitzer} linear motions and we already have MOPEX mosaics/extractions in hand for each target's field ($\S$\ref{sec:spitzer_photometry}), we opt to perform a mosaic-based \textit{Spitzer} astrometric analysis.

Our approach for deriving recalibrated \textit{Spitzer} ch2 astrometric measurements from the p14034 imaging is largely modeled after the procedure laid out in \cite{martin_spitzer_astrometry}. Our starting point is the list of ch2 MOPEX/APEX pixel coordinate centroids and (RA, Dec) world coordinate values for all sources in each of our AORs. These are the same MOPEX/APEX catalogs used previously to obtain our ch2 magnitudes for each target in $\S$\ref{sec:spitzer_photometry}. The world coordinates natively provided by MOPEX/APEX rely on a WCS solution calibrated to 2MASS, and typically show offsets of several tenths of a mosaic pixel (0.6$''$/pixel) relative to \textit{Gaia}. The systematics inherent in the initial 2MASS WCS solutions are thus much larger than the systematics floor attainable through astrometric recalibration to \textit{Gaia}.

\subsubsection{ch2 Astrometric Recalibration}
\label{sec:spitzer_astrom_recalib}

The primary challenge in recalibrating each ch2 mosaic's astrometry is identifying a sufficient number of \textit{Gaia} DR2 calibration sources ($N_{calib}$) with high-SNR ch2 counterparts. In selecting astrometric calibrators, we always restrict to the `non-flanking' region of each p14034 ch2 mosaic (the portion of the mosaic built from single-frame BCD images which contain the brown dwarf candidate's location). Given our dither strategy, this results in a full frame coverage sky area from which to select \textit{Gaia} calibrators of only 13-15 square arcminutes, depending on the number of dithers. 

Ideally, we would be able to obtain at least 10 \textit{Gaia} DR2 calibrators in each ch2 mosaic's non-flanking, full-coverage sky region with very high SNR ch2 counterparts (SNR $\ge$ 100) and \textit{Gaia} proper motions available. However, this was possible for just 67 of our 164 ch2 mosaics (see the first row of Table \ref{tab:spitzer_recalib_samples}). In cases where this ideal set of calibrator selection criteria (which we refer to as method \#1) yielded fewer than 10 calibrators, we tried a sequence of somewhat loosened cuts in order to always obtain at least 5 \textit{Gaia} calibrators\footnote{5 is the minimum number of calibrators used for any field in the \textit{Spitzer} ch2 parallax-fitting astrometric analysis of \cite{davy_parallaxes}.}. We sequentially tried each set of selection criteria listed in Table \ref{tab:spitzer_recalib_samples}, in order of ascending ``method number'' (leftmost column in Table \ref{tab:spitzer_recalib_samples}) until a set of selection cuts yielded $N_{calib} \ge N_{calib, min}$. Method \#2 is the same as our ideal set of cuts, but reduces the ch2 counterpart SNR threshold to 50. Method \#3 further reduces the ch2 SNR threshold to 30 and additionally reduces the minimum required number of \textit{Gaia} calibrators from 10 to 5 (as specified in the $N_{calib, min}$ column of Table \ref{tab:spitzer_recalib_samples}).

Methods \#4, \#5, \#6 are the same as methods \#1, \#2, \#3 respectively, but with the requirement of \textit{Gaia} DR2 proper motion availability dropped (see boolean `\textit{Gaia} PM required' column of Table \ref{tab:spitzer_recalib_samples}). Our inability to correct for calibrator motion between the \textit{Gaia} 2015.5 epoch and our \textit{Spitzer} ch2 epoch a few years later is regrettable but a necessary compromise  when resorting to methods \#4-\#6. The typical motions of our \textit{Gaia} DR2 calibrators are very small, so this should have a negligible impact on our final WISE+\textit{Spitzer} linear motion results, and only 7 of 164 AORs end up using \textit{Gaia} calibrators lacking proper motions.

Methods \#7, \#8, \#9 are the same as methods \#1, \#2, \#3 respectively, but reduce the minimum mosaic BCD frame coverage requirement from full coverage to at least 50\% coverage. While this is not ideal, only 5 of 164 AORs needed to employ this lowered frame coverage requirement.

The `method number' column of Table \ref{tab:spitzer_positions} specifies which set of \textit{Gaia} DR2 calibrator selection criteria was used in determining each row's recalibrated \textit{Spitzer} ch2 position. The $N_{calib}$ column of Table \ref{tab:spitzer_positions} lists the number of \textit{Gaia} DR2 calibrators employed for each recalibrated ch2 position measurement. The minimum (maximum) number of astrometric calibrators per AOR is 5 (92). The median (mean) number of astrometric calibrators per AOR is 12 (16). In all cases where our selection criteria demand that \textit{Gaia} DR2 calibrators have proper motions available, we use these proper motions to propagate the \textit{Gaia} calibrator positions to the \textit{Spitzer} ch2 epoch.

Having selected a set of \textit{Gaia} DR2 calibrators for each AOR, we proceed to re-fit 6 parameters of each ch2 mosaic's WCS: all four elements of the CD matrix and the two CRPIX components. The bright end systematics floor achieved via our recalibrated ch2 mosaic WCS solutions is very similar to that of \cite{martin_spitzer_astrometry}. The median per-mosaic bright end scatter is 25 (23) mas in RA (Dec). The 16th-84th percentile ranges are 15-44 mas in RA and 14-39 mas in Dec. For comparison, \cite{martin_spitzer_astrometry} cites a typical
systematics floor of $\sim$15-40 mas for their ch2 mosaic WCS recalibration.

Table \ref{tab:spitzer_positions} lists the recalibrated ch2 (RA, Dec) position obtained for each target, including metadata such as the AOR used and the MJD. The uncertainties $\sigma_{RA}$ and $\sigma_{Dec}$ are computed by summing the per-AOR bright end scatter in quadrature with the statistical uncertainty on each target's ch2 centroid measurement.

\subsection{WISE+\textit{Spitzer} Linear Motion Fits}
\label{sec:poly_fit}

For each brown dwarf candidate, we gather its combined list of WISE and \textit{Spitzer} positions, corresponding MJD values, and positional uncertainties from Tables \ref{tab:wise_positions} and \ref{tab:spitzer_positions}. The typical combined WISE+\textit{Spitzer} time baseline for targets with p14034 imaging available is $\sim$8.6 years. Along each coordinate direction (RA and Dec) we fit a linear model to the combined list of WISE and \textit{Spitzer} positions as a function of MJD in order to measure $\mu_{\alpha}$ and $\mu_{\delta}$. Throughout this paper the quoted $\mu_{\alpha}$ values are in angular rather than coordinate units i.e., they already have the cos($\delta$) factor multiplied into them. Through these same per-coordinate linear fits we also obtain parameters $\alpha_0$ and $\delta_0$, the object's (RA, Dec) coordinates at a fiducial time MJD$_{0}$. MJD$_0$ is the inverse variance weighted mean MJD of the contributing astrometric data points. MJD$_{0}$ ends up typically being similar to the MJD of the \textit{Spitzer} observation, since the ch2 positional uncertainties are much smaller than those in W2. Our linear fits are performed using weighted linear least squares and so naturally produce uncertainties on the fiducial location and best-fit linear motion components via simple matrix algebra. The measurement uncertainties fed to the weighted linear least squared routine are the $\sigma_{RA}$ and $\sigma_{Dec}$ values provided in Tables \ref{tab:wise_positions} and \ref{tab:spitzer_positions}. No rescaling of the $\sigma_{RA}$, $\sigma_{Dec}$ positional uncertainties is performed.

We do not allow for any outlier rejection in our per-coordinate linear motion fits. All detections used in our WISE+\textit{Spitzer} motion fits were visually vetted, so there should be no need for outlier rejection.

For our targets which turned out to have two \textit{Spitzer} counterparts ($\S$\ref{sec:spitzer_pairs}), we performed separate linear motion fits for each \textit{Spitzer} counterpart, where the two motion fits both use the same set of WISE detections (since in such cases the brown dwarf candidate appears as just a single object in WISE).

Table \ref{tab:wise_spitzer_pm} lists our linear motion fit results for all targets. The $\mu_{\alpha}$, $\sigma_{\alpha_0}$ and $\sigma_{\mu_{\alpha}}$ values are all in angular rather than coordinate units. The total motion $\mu_{tot}$ values are calculated by summing the RA and Dec linear motion components in quadrature, and the quoted $\mu_{tot}$ uncertainties are based on first order propagation of the $\sigma_{\mu_{\alpha}}$, $\sigma_{\mu_{\delta}}$ errors.

\begin{figure}
\plotone{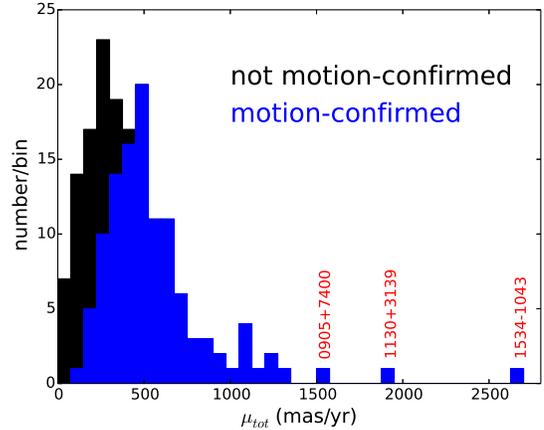}
\caption{Histogram of total motion measurements resulting from our sample's linear motion fits ($\S$\ref{sec:poly_fit}). The number of targets per bin which we could not motion confirm is shown in black, stacked on top of the number of motion-confirmed targets per bin (blue). We exclude the 6 spurious candidates (Table \ref{tab:duds}), 4 CPM objects, and the two targets which each have two \textit{Spitzer} counterparts ($\S$\ref{sec:spitzer_pairs}). Three of our motion-confirmed discoveries have best-fit $\mu_{tot} > 1500$ mas/yr, and twelve have $\mu_{tot} > 1000$ mas/yr.}
\label{fig:total_motion_hist}
\end{figure}

Figure \ref{fig:total_motion_hist} shows a histogram that vertically stacks the number of non motion-confirmed targets (black) on top of the number of motion-confirmed targets (blue). Targets with $\mu_{tot} > 380$ mas/yr are always motion confirmed. The motion-confirmed fraction decreases to 50\% at $\mu_{tot} \approx 265$ mas/yr. The minimum best-fit $\mu_{tot}$ of any motion-confirmed member of our sample is 139 mas/yr. The minimum value of $\mu_{tot}/\sigma_{\mu_{tot}}$ for any motion-confirmed target is 4.85, which suggests that, roughly speaking, our $\chi^2_{motion} = 23.03$ threshold is akin to a requirement of 5$\sigma$ significant motion. 114 of our discoveries are motion confirmed via the astrometric analysis presented in this work, as indicated in the boolean `motion confirmed' column of Table \ref{tab:wise_spitzer_photom}.

\section{Ground-based Photometry}
\label{sec:ground_based}

Although \textit{Spitzer} ch1$-$ch2 color provides the most efficient means for constraining the spectral types of our brown dwarf candidates, $J$ band photometry can further inform spectral type estimates and represents a crucial step toward ultimately obtaining NIR spectroscopic confirmations. Beyond T5, $J-$W2 color increases rapidly toward later spectral types, with the T/Y boundary at $J-$W2 $\sim$ 5 mag \citep{kirkpatrick11}. For candidates thought to be potential very late T or Y dwarfs, we therefore sought to obtain ground-based $J$ band follow up to a depth of at least $J \sim 21$ (given that W2 $\approx$ 16 is typical for our sample as discussed in $\S$\ref{sec:sample_properties}). Our NIR imaging was not intended to be used for high-fidelity astrometry, and we do not attempt to incorporate ground-based NIR data into our astrometric analyses for several reasons. Among these, the NIR observations generally do not extend our overall time baseline appreciably beyond our \textit{Spitzer} p14034 epoch.

\subsection{Gemini/FLAMINGOS-2 Follow-up}

For southern targets, we acquired follow-up $J$ band imaging at Gemini South. Through program GS-2019A-Q-316 (PI Gelino) we obtained FLAMINGOS-2 \citep{flamingos2} $J$ band photometry for 16 of the discoveries presented in this paper. The $J$ band photometry from Gemini that we provide is in the MKO photometric system. To calibrate the photometry in each target's field using 2MASS, we applied the color transformation equation from \cite{hodgkin_2009}:

\begin{equation}
\label{equ:mko_calib}
J_{MKO} = J_{2MASS} - 0.065 \times (J_{2MASS} - H_{2MASS})
\end{equation}

For each target, we requested a 60 second exposure time per dither with a 9-position dither pattern.

\subsection{Palomar Hale/WIRC Follow-up}

For northern targets, we obtained follow-up $J$ band imaging with the Wide Field Infrared Camera \citep[WIRC;][]{WIRC} at the Palomar Hale 200 inch telescope (PI Marocco). This WIRC photometry is also in the MKO system, and we again used Equation \ref{equ:mko_calib} to photometrically calibrate each field. We obtained WIRC $J$ band photometry for 32 of the discoveries presented in this paper. The total exposure time varied depending on environmental conditions and the anticipated $J$ band magnitude of each target, but 15 dithers at 2 minutes per dither represents a typical observing sequence for one object.

\subsection{Archival Near-infrared Photometry}
\label{sec:archival_nir}

\subsubsection{2MASS}
\label{sec:2mass}

By selection, we expect very few of our targets to have 2MASS counterparts. Nevertheless, we checked our entire sample for 2MASS counterparts. We used our best-fit WISE+\textit{Spitzer} proper motions from Table \ref{tab:wise_spitzer_pm} to predict each target's (RA, Dec) at a fiducial 2MASS epoch of year = 1999.5, then visually inspected 2MASS images to look for a counterpart at that location. In all, we only found 2MASS counterparts\footnote{In both cases the 2MASS counterpart is drawn from the Point Source Reject Table.} to two discoveries in our sample: CWISEP 1402+1021 and CWISEP 2015$-$6750. Both of these are relatively bright/blue members of our sample, selected specifically because we expected they may be mid-T brown dwarfs potentially within a distance of 20 pc. Both 2MASS counterparts are detected only in $J_{2MASS}$, and in both cases the $J_{2MASS}$ magnitude is in good agreement with that predicted based on ch2 magnitude and \textit{Spitzer} phototype. Table \ref{tab:tmass} lists the two 2MASS counterparts recovered. The magnitude limits in Table \ref{tab:tmass} are based on 95\% confidence flux upper limits.

\subsubsection{UKIRT/WFCAM and VISTA/VIRCAM}
\label{sec:uhs_vhs}

We searched the entire WFCAM Science Archive (WSA) and VISTA Science Archive (VSA) for $JHK/K_S$ counterparts to our brown dwarf candidates. Specifically, we searched all ``pawprint'' exposure sets\footnote{\url{http://www.vista.ac.uk/glossary.htm\#pawprint}} through 2017 January 1 in VSA and 2014 March 7 in WSA\footnote{Public availability of data in VSA/WSA becomes a relatively complex issue at later dates.}. We queried for counterparts within a 5$''$ radius of a nominal epoch $\sim$2014 position for each target, based on our linear motion solutions. We retrieved all matched VSA/WSA detections within this relatively large 5$''$ radius, allowing us to perform more detailed disambiguation downstream. Further, when possible, we retrieved available $5\sigma$ magnitude limits in cases where no matches were found within a $5''$ radius yet archival VSA/WSA imaging at the target location exists.

In total we retrieved $\sim$1,100 VSA/WSA matched detections or magnitude limits in $JHK/K_{S}$. These were drawn predominantly from the UKIRT Hemisphere Survey \citep[UHS;][]{uhs} and VISTA Hemisphere Survey \citep[VHS;][]{vhs}, but also incorporate contributions from a variety of other smaller-area surveys including ULAS \citep{ulas}, VIKING \citep{viking}, and a few PI programs. In many cases a single brown dwarf candidate has multiple VSA/WSA detections/limits in one NIR band.

We therefore sought to condense/vet our raw VSA/WSA query results and thereby compose a summary consisting of at most one VSA or WSA magnitude or magnitude limit per NIR band per target. To do so, we used the linear motion solutions from Table \ref{tab:wise_spitzer_pm} to predict the moving target's position at each VSA/WSA pawprint MJD. We then retained only matches within 1.5$''$ of each predicted position. In cases where a single target has multiple matched detections within 1.5$''$ in a single band, we adopt the magnitude of the closest match. When a target has no counterparts in a given band within 1.5$''$, we quote a magnitude limit in that band if one is available. In cases where there is no counterpart but multiple limits, we adopt the deepest limit. We also enforced a veto list containing a small handful of VSA/WSA detections which were noted to be static contaminants (wrong cross-matches) based on visual inspection of the VIRCAM/WFCAM imaging. We additionally discarded all VSA/WSA detections with nonzero PPERRBITS data quality flags\footnote{\url{http://wsa.roe.ac.uk/ppErrBits.html}} and/or with an extended morphological classification (CLASS = 1), so as to avoid artifacts and incorrectly matched galaxies.

Photometry from WFCAM $JHK$ and VIRCAM $JH$ is in the MKO system, whereas VIRCAM employs a $K_{S}$ filter.

\subsubsection{Merging Archival \& Follow-up NIR Photometry}

\begin{figure*}
\plotone{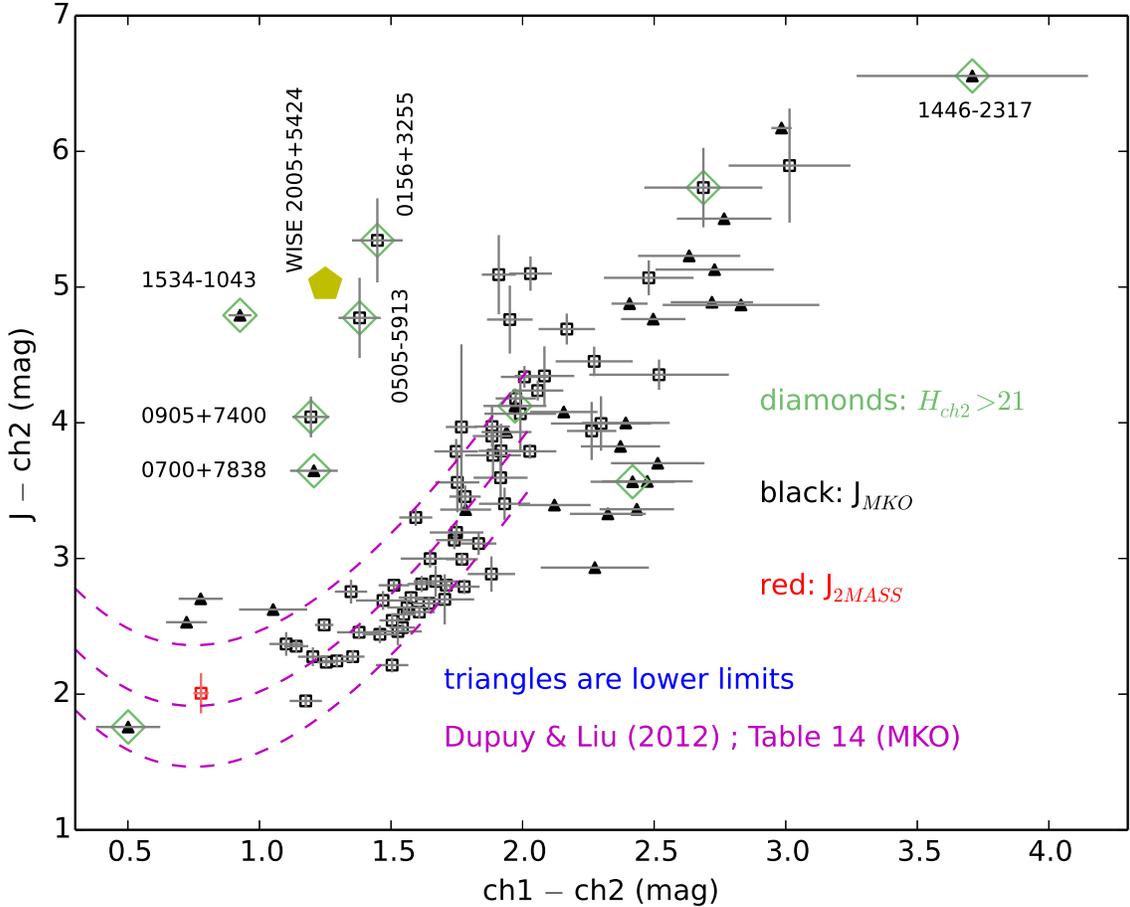}
\caption{$J - $ch2 versus ch1$-$ch2 for motion-confirmed discoveries with \textit{Spitzer} photometry available. The $J$ band photometry is drawn from our Gemini/FLAMINGOS-2 and Palomar/WIRC follow-up, the WFCAM and VISTA archives, and 2MASS. The purple dashed lines trace the expected trend for late-type brown dwarfs from \cite{dupuy_liu_2012}, plus/minus the scatter in the relevant relations from their Table 14. Our motion-confirmed brown dwarf candidates largely follow this trend while continuing (as expected) to become even redder in $J - $ch2 beyond T9, where the \cite{dupuy_liu_2012} relations are not intended to be applicable. Light green diamonds indicate targets with exceptionally large reduced proper motion, $H_{ch2} > 21$ mag. Five targets (each specially labeled with its short name) are anomalously red in $J - $ch2 relative to their ch1$-$ch2 colors, inhabiting a region of parameter space with $J - $ch2 $ > 3.5$ and ch1$-$ch2 $ < 1.5$. All five of these major color outliers have $H_{ch2} > 21$, suggesting that they may represent a low metallicity subpopulation. Color limits are based on $5\sigma$ $J$ band flux limits. The yellow pentagon denotes the location of the benchmark T8 subdwarf WISE 2005+5424 within this color-color space.}
\label{fig:j_minus_ch2}
\end{figure*}

\begin{figure*}
\plotone{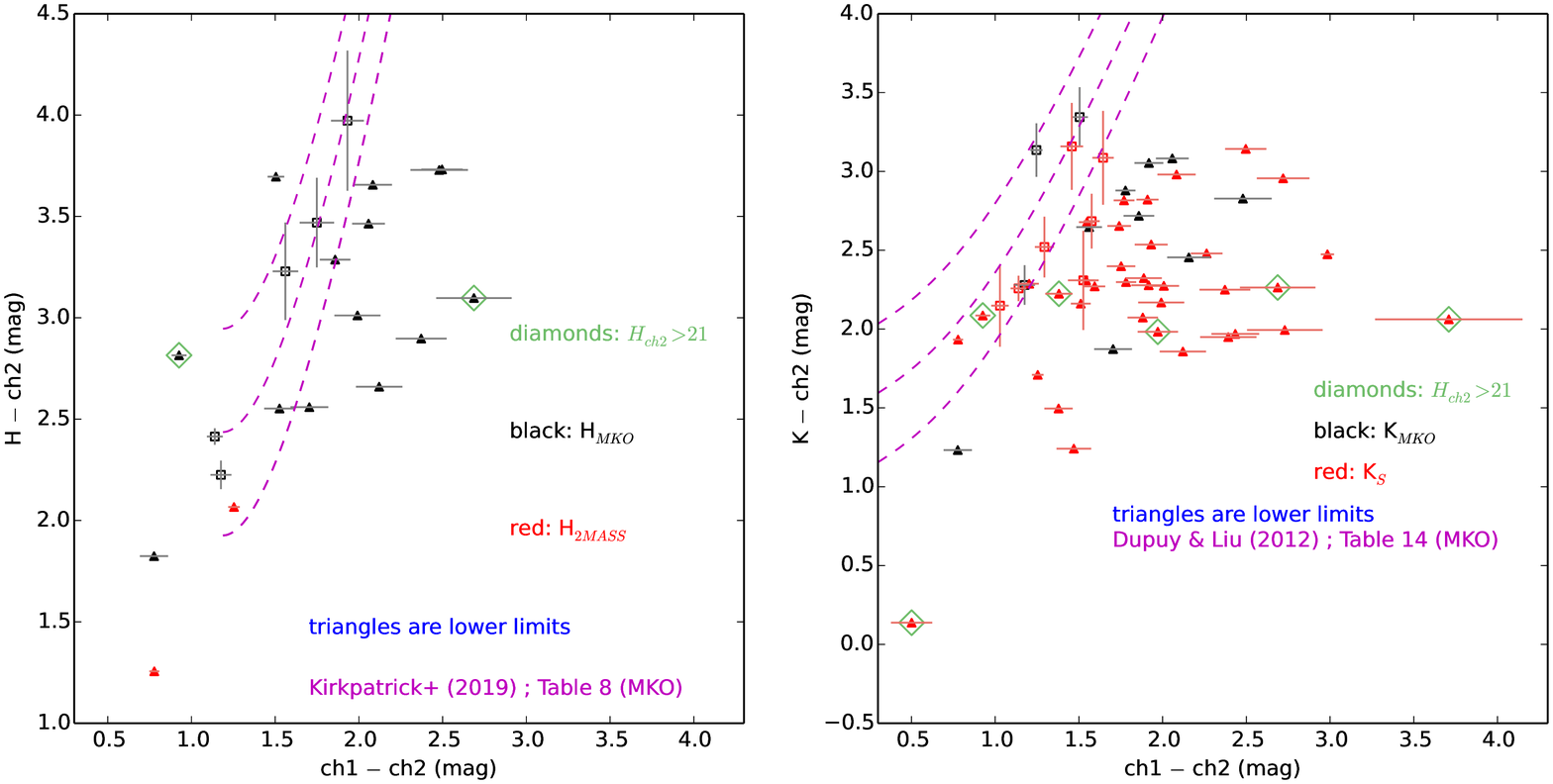}
\caption{Same as Figure \ref{fig:j_minus_ch2}, but for $H - $ch2 (left) and $K - $ch2 (right). Relative to $J$, there are far fewer targets with available data at $H$ and $K$, in part reflecting the availability of VISTA/UKIRT archival observations, but also because our NIR imaging follow-up campaigns at Gemini and Palomar only employ $J$ band. The limited number of detections follow the literature trends reasonably well. In each panel, the CWISEP 1402+1021 and CWISEP 2015$-$6750 color lower limits are based on 95\% confidence flux upper limits from 2MASS. All other color lower limits are based on $5\sigma$ flux upper limits.}
\label{fig:hk_minus_ch2}
\end{figure*}

In order to produce color-color diagrams such as those of Figure \ref{fig:j_minus_ch2} and Figure \ref{fig:hk_minus_ch2}, we sought to merge our Palomar, Gemini, WFCAM, VIRCAM and 2MASS NIR photometry into a single compilation with at most one magnitude or magnitude limit per NIR band per target. In doing so, we always give precedence to our dedicated CatWISE follow-up over archival information in $J$ band when both options are available. In one isolated instance, we have archival photometry available from both VISTA and 2MASS while lacking CatWISE follow-up: $J$ band for CWISEP 2015$-$6750. In this case we adopt the much higher SNR measurement from VISTA. Table \ref{tab:jhk_phot} provides the merged compilation of VSA/WSA and follow-up $JHK/K_S$ magnitudes and limits for motion-confirmed discoveries with at least one such NIR magnitude or limit available. 2MASS photometry is listed separately in Table \ref{tab:tmass}. All limits quoted in Table \ref{tab:jhk_phot} are $5\sigma$. Finally, Table \ref{tab:j_follow_up_no_motion} lists our Palomar and Gemini $J$ band follow-up for sources that were not motion-confirmed by our astrometric analysis.

\section{Discussion}
\label{sec:discussion}

\subsection{Photometric Spectral Type Estimates}
\label{sec:phototypes}

Obtaining photometric spectral type estimates was the primary motivation for conducting our \textit{Spitzer} p14034 follow-up campaign. We use the ch1$-$ch2 colors from Table \ref{tab:wise_spitzer_photom} to estimate spectral types. We do not attempt to fold NIR magnitudes/limits into our phototyping in this work. For our spectral type estimates, we use a type versus ch1$-$ch2 color grid constructed from the \cite{davy_parallaxes} relation for T5.5-Y1 and \cite{kirkpatrick11} for T0-T5. These grids are quantized at the level of 0.5 type. Quoted spectral type estimates result from evaluating these type versus ch1$-$ch2 grids at the best fit (i.e., central) ch1$-$ch2 color from Table \ref{tab:wise_spitzer_photom}. Our photometric type estimates do not extend colder than Y1 due to the scarcity of empirical data in this regime, and as a result objects with best-fit phototypes of Y1 are listed as $\geq $Y1 in Tables \ref{tab:derived_properties} and \ref{tab:possible_y_dwarfs}. Table \ref{tab:derived_properties} presents a number of derived properties for each of our motion-confirmed discoveries, including our spectral type estimates in a column labeled SpT for short. 

\begin{figure}
\plotone{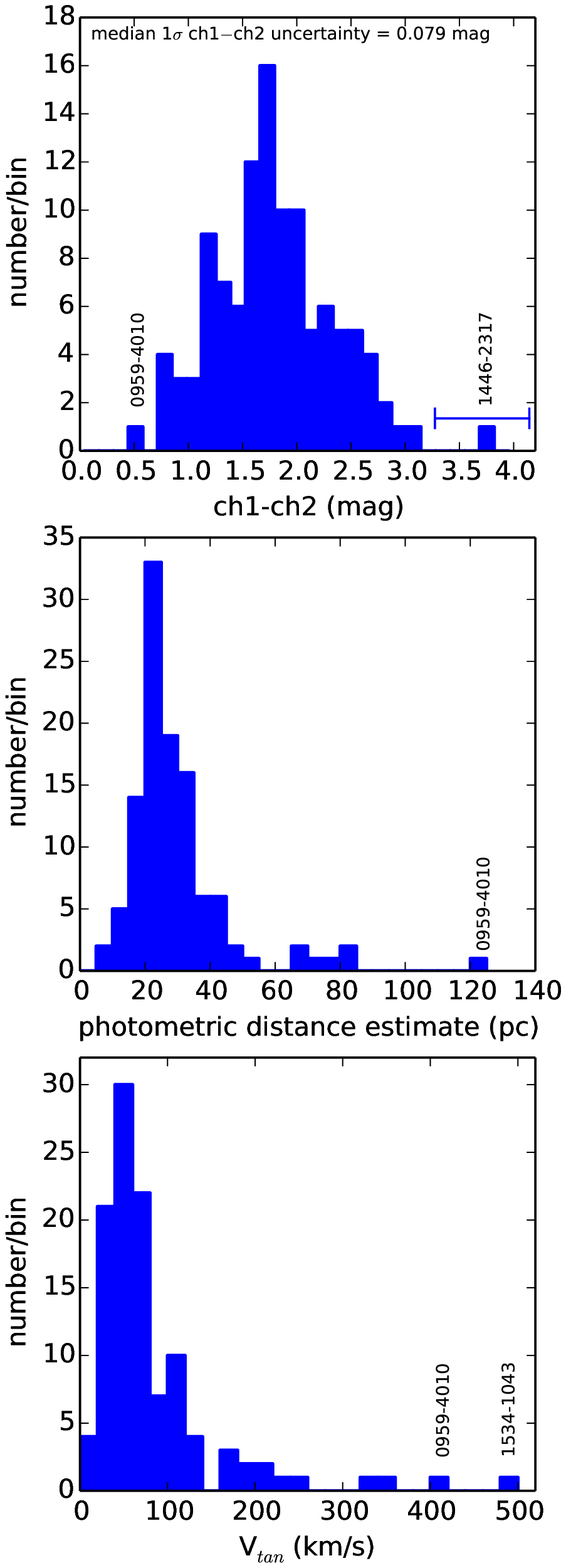}
\caption{Histograms of measured \textit{Spitzer} colors (top) and selected derived parameters (middle: photometric distance estimates, bottom: $V_{tan}$ estimates) from Table \ref{tab:derived_properties} for motion-confirmed discoveries with \textit{Spitzer} photometry available. Short names of outliers are added as annotations. As discussed in $\S$\ref{sec:color_color}, we hypothesize that several objects with unusually large $V_{tan}$ and/or distance estimates may be late-type subdwarfs.}
\label{fig:gold_sample_hist}
\end{figure}

Figure \ref{fig:gold_sample_hist} shows a histogram of our measured ch1$-$ch2 colors for motion-confirmed brown dwarf candidates. The median ch1$-$ch2 color within our motion-confirmed sample is 1.75 mag, corresponding to a spectral type of approximately T8. CWISEP 0959$-$4010 has the bluest ch1$-$ch2 color of any motion-confirmed target in our sample, ch1$-$ch2 = $0.50 \pm 0.12$ mag, for which we obtain a spectral type estimate of T3.5. The reddest best-fit ch1$-$ch2 color of any motion-confirmed target in our sample is $3.71 \pm 0.44$ mag for CWISEP 1446$-$2317, which has its reported spectral type estimate listed as $\geq$Y1.

We caution against overinterpretation of our spectral type estimates on an object-by-object basis, since we regard these as considerably uncertain for a number of reasons. The \cite{davy_parallaxes} tabulation of ch1$-$ch2 versus spectral type does not provide a direct formula for spectral type as a function of  ch1$-$ch2 color, nor a prescription for quoting uncertainties on spectral type estimates inferred from \textit{Spitzer} color. Additionally, spectral typing becomes relatively poorly defined at types $\gtrsim $Y1, due to a number of factors including small sample size, lack of NIR spectroscopic data \citep[e.g., WISE 0855$-$0714, WD  0806$-$661 B;][]{wd0806}, and difficulty fitting all examples into a common sequence of spectral morphology \citep[see WISE 1828+2650;][]{beichman_w1828, leggett_w1828}.

Although these considerations lead us not to quote per-target spectral type errors, we can still use the \cite{davy_parallaxes} compilation of spectral types and ch1$-$ch2 colors to provide an overall sense for the level of uncertainty on our type estimates. We fit a second order polynomial to the (ch1$-$ch2, spectral type) pairs with 0.9 $\le$ ch1$-$ch2 $\le$ 3.0 in the bottom center panel of \cite{davy_parallaxes} Figure 4, using ch1$-$ch2 as the independent variable. The residuals relative to the best fit show an RMS scatter of $\pm$0.56 in spectral type.  As a result of the p14034 observing strategy, our ch1$-$ch2 color uncertainties ramp up from $\sim$0.06 mag for ch1$-$ch2 $ < $ 1.2 to $\sim$0.2 mag at type $\sim$Y1. Propagating this color uncertainty through our best-fit polynomial relation for spectral type, this translates into a scatter of $\sim$0.15 ($\sim$0.65) in type at mid-T (Y1). Adding the worst-case 0.65 type scatter in quadrature with that of the type versus ch1$-$ch2 polynomial residuals and that from our 0.5 type quantization yields an overall errorbar of roughly 1.0 in spectral type for our reported estimates.

We emphasize that it is not actually possible to measure a spectral type with the data at hand --- our ch1$-$ch2 colors are measured in a totally different wavelength regime than that in which the spectral type is defined (at $J$ and $H$ bands), so we can only provide our best estimate as to near-IR type based on its known correlation with \textit{Spitzer} color. True spectral types can only be measured using spectra in the wavelength range of interest. In the absence of observationally prohibitive spectroscopic confirmations, spectral type estimation for our coldest motion-confirmed discoveries would greatly benefit from further astrometric follow-up in the future, with trigonometric distances yielding critical ch2 absolute magnitude estimates.

\subsection{Photometric Absolute Magnitude and Distance Estimates}
\label{sec:abs_mags_dist}

We use the \citet[][Table 8]{davy_parallaxes} ch1$-$ch2 to $M_{ch2}$ relation to obtain photometric absolute magnitude estimates, which in turn yield corresponding photometric distance estimates. These derived properties for our motion-confirmed targets are presented in Table \ref{tab:derived_properties}. When calculating uncertainties on these absolute magnitude estimates, we take into account both our $1\sigma$ ch1$-$ch2 color uncertainties and the 0.3 mag residual scatter of the $M_{ch2}($ch1$-$ch2) polynomial fit relative to its training data.

The \cite{davy_parallaxes} $M_{ch2}($ch1$-$ch2) relation is only valid for $0.9 \le $ch1$-$ch2$ \le 3.7$. However, a handful of our motion-confirmed discoveries have ch1$-$ch2 $ < 0.9$ mag, and in these cases we obtain $M_{ch2}$ estimates by plugging our spectral type estimates from Table \ref{tab:derived_properties} into the \citet[][Table 14]{dupuy_liu_2012} $M_{ch2}$(SpT) relation, which is applicable in the relevant early-mid T regime.

CWISEP 1446$-$2317 is our only discovery with best-fit ch1$-$ch2 color too red for the \cite{davy_parallaxes} $M_{ch2}$(ch1$-$ch2) relation. In this case, we quote only a lower limit of $M_{ch2} > 16.23$. This limit is derived by applying the \cite{davy_parallaxes} $M_{ch2}$(ch1$-$ch2) relation to the ch1$-$ch2 color obtained by subtracting this object's 1$\sigma$ \textit{Spitzer} color uncertainty from its central ch1$-$ch2 color. Correspondingly, we quote only a distance upper limit of 8.3 pc for CWISEP 1446$-$2317 based on its $M_{ch2}$ lower limit. Such limits should be treated with caution given our poor constraint on this object's \textit{Spitzer} color.

The middle panel of Figure \ref{fig:gold_sample_hist} shows a histogram of our motion-confirmed sample's photometric distance estimates. As discussed in $\S$\ref{sec:color_color}, we suspect that many of the objects with unusually large distance estimates ($\gtrsim$50 pc) may be subdwarfs. The median distance estimate for motion-confirmed targets is 26 pc. Propagating our $M_{ch2}$ estimates and their errorbars into distance estimates, the $1\sigma$ fractional uncertainty on our photometric distances is typically $\sim$15\%.

\subsection{Photometric Effective Temperature Estimates}
\label{sec:teff}

Table \ref{tab:derived_properties} also provides $T_{eff}$ estimates based on the \citet[][Table 8]{davy_parallaxes} $T_{eff}$(ch1$-$ch2) polynomial relation. When calculating uncertainties on these $T_{eff}$ estimates, we take into account both our $1\sigma$ ch1$-$ch2 color uncertainties and the 81 Kelvin residual scatter of the $T_{eff}($ch1$-$ch2) polynomial fit relative to its training data. This floors our quoted $T_{eff}$ uncertainties at 81 Kelvin.

The \cite{davy_parallaxes} $T_{eff}($ch1$-$ch2) relation is only valid for $0.9 \le $ch1$-$ch2$ \le 3.7$. We omit $T_{eff}$ estimates for the handful of our motion-confirmed discoveries with ch1$-$ch2 $ < $ 0.9 mag. There is currently no $T_{eff}($ch1$-$ch2) relation available in this early-mid T regime, and within this range of spectral types $T_{eff}$ maintains a roughly uniform value of $\sim$1400-1500 Kelvin anyway \citep[e.g.,][Figure 7]{kirkpatrick_review_2005}.

The best-fit CWISEP 1446$-$2317 \textit{Spitzer} color is too red for the \cite{davy_parallaxes} $T_{eff}($ch1$-$ch2) relation. We therefore quote only an upper limit of $T_{eff} < 381$ K for CWISEP 1446$-$2317. This value arises from evaluating the $T_{eff}($ch1$-$ch2) formula at a color $1\sigma$ bluer than the central value and further adding the aforementioned 81 K systematics floor of the $T_{eff}$ relation in attempt to be conservative.

\subsection{Reduced Proper Motions and Tangential Velocities}
\label{sec:rpm_vtan}

\begin{figure}
\plotone{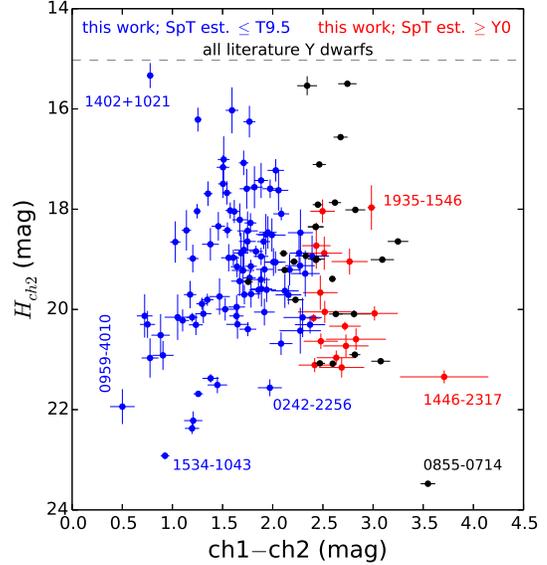}
\caption{Reduced proper motion diagram showing all Y dwarfs from the prior literature (black) and all motion-confirmed discoveries from this study with \textit{Spitzer} photometry available. Our targets with best-fit ch1$-$ch2 color most consistent with spectral type Y are shown in red, while all of our discoveries with earlier spectral type estimates are shown in blue.}
\label{fig:reduced_pm}
\end{figure}

Reduced proper motions ($H_{ch2}$) can be calculated entirely on the basis of directly measured quantities: $\mu_{tot}$ (Table \ref{tab:wise_spitzer_pm}) and apparent ch2 magnitude (Table \ref{tab:wise_spitzer_photom}). In calculating uncertainties on $H_{ch2}$ we account for the uncertainties on both $\mu_{tot}$ and ch2 magnitude. Table \ref{tab:derived_properties} lists $H_{ch2}$ for all motion-confirmed targets with \textit{Spitzer} imaging available. Figure \ref{fig:reduced_pm} shows a scatter plot of $H_{ch2}$ versus ch1$-$ch2 color for the same set of our discoveries, plus all Y dwarfs from the prior literature to provide context. Interestingly, all three of our motion-confirmed discoveries with $H_{ch2} > 22$ mag are rather blue in ch1$-$ch2 color relative to the bulk of our sample. These three objects (CWISEP 0700+7838, CWISEP 0905+7400, WISEA 1534$-$1043) have ch1$-$ch2 $ < $1.25 mag, making all of them bluer in \textit{Spitzer} color than at least 85\% of our motion-confirmed sample.

Computing $V_{tan}$ values requires making use of our photometric distance estimates from $\S$\ref{sec:abs_mags_dist}. In computing uncertainties on $V_{tan}$ we account for the uncertainties on both $\mu_{tot}$ and our distance estimates. Table \ref{tab:derived_properties} provides our $V_{tan}$ estimates for motion-confirmed targets with \textit{Spitzer} imaging available. The bottom panel of Figure \ref{fig:gold_sample_hist} displays a histogram of the central $V_{tan}$ values for this sample. As discussed in $\S$\ref{sec:color_color}, a number of our highest $V_{tan}$ motion-confirmed discoveries are also color outliers.

For CWISEP 1446$-$2317, our distance upper limit from Table \ref{tab:derived_properties} translates into a $V_{tan}$ upper limit, $V_{tan} < 53$ km/s.

\subsection{Y Dwarf Candidates}
\label{sec:likely_y_dwarfs}

Seventeen of our motion-confirmed discoveries have best-fit phototypes $\geq$Y0. In practice, within the spectral type estimation framework described in $\S$\ref{sec:phototypes}, this is to say that seventeen of our motion-confirmed targets have central ch1$-$ch2 colors of at least 2.40 mag. Plots of our astrometric measurements and linear motion fits for all 17 of these Y dwarf candidates are shown in Figure \ref{fig:radec_plots}. An eighteenth target, CWISEP 0212+0531, has ch1$-$ch2 $> 2.40$ mag, but its motion appears to be disconfirmed (or else very small) in light of our combined WISE and \textit{Spitzer} astrometry.

Given our $\S$\ref{sec:phototypes} appraisal that SpT estimates for our reddest objects carry an uncertainty of roughly 1 type, we cannot yet specify an exact list of our targets that will ultimately turn out to fall on the late-type side of the T/Y boundary. Nevertheless, we can perform some cross-checks on the hypothesis that our 17 reddest motion-confirmed discoveries are indeed Y dwarfs, and rate the relative likelihood that particular objects in this sample are/aren't in fact `merely' T dwarfs.

We find that the archival and follow-up $JHK/K_S$ NIR photometry assembled in $\S$\ref{sec:ground_based} is  consistent with the proposition that most motion-confirmed objects with ch1$-$ch2 $\geq$ 2.4 mag are Y dwarfs. All 17 targets with phototypes $\geq$Y0 have $J$ band imaging available. Four of these targets have $J$ band detections: CWISEP 0938+0634,  CWISEP 2256+4002, CWISEP 2355+3804 and CWISEP 2356$-$4814. Their respective $J - $ch2 colors are $5.07 \pm 0.12$, $5.90 \pm 0.41$, $4.35 \pm 0.10$ and $5.73 \pm 0.28$ magnitudes. Three of these $J - $ch2 colors are consistent with Y spectral types. The fourth $J - $ch2 color, for CWISEP 2355+3804, seems to provide an indication that this target may actually be a late T dwarf.

The remaining 13 targets with type estimates $\geq$Y0 have $J - $ch2 limits available. Among these, five objects have $5\sigma$ $J$ limits establishing that $J - $ch2 $> 5$ mag,  bolstering the case (though still not guaranteeing) that they are Y dwarfs: CWISEP 0238$-$1332 ($J - $ch2 $> 5.13$), CWISEP 0321+6932 ($J - $ch2 $> 5.23$), CWISEP 0940+5233 ($J - $ch2 $> 5.50$), CWISEP 1446$-$2317 ($J - $ch2 $> 6.56$) and CWISEP 1935$-$1546 ($J - $ch2 $> 6.17$). Another 4 motion-confirmed candidates with ch1$-$ch2 $ \geq $ 2.40 mag have $J - $ch2 limits corresponding to minimum colors that would be extremely red among the latest T dwarfs: CWISEP 0402$-$2651 ($J -$ ch2$ > 4.89$), WISENF 1936+0408 ($J - $ ch2$ > 4.88$), CWISEP 2011$-$4812 ($J - $ ch2$ > 4.76$) and CWISEP 2230+2549 ($J - $ ch2 $ > 4.87$). Lastly, four targets with type estimates $\geq$Y0 have $J$ nondetections establishing relatively weak $J - $ch2 limits in the 3.35-3.70 mag range (CWISEP 0634+5049, CWISEP 0859+5349, CWISEP 1047+5457 and  CWISEP 1359$-$4352). Clearly, deeper $J$ band imaging follow-up would be highly valuable in pinning down which side of the T/Y boundary each of our $\geq$Y0 phototype discoveries falls on.

None of our $\geq$Y0 phototype targets have detections at $H$ band. Three of our $\geq$Y0 phototype targets have $H$ nondetections establishing $H - $ch2 limits in the range of 3.1-3.75 mags (CWISEP 0938+0634, CWISEP 2011$-$4812 and CWISEP 2356$-$4814). While these color limits are consistent with spectral type Y, they are essentially uninformative --- an $H - $ch2 limit would need to reach $\gtrsim$5 mag in order to begin disfavoring late T spectral types \citep[e.g.,][Figure 4]{davy_parallaxes}.

None of our $\geq$Y0 phototype targets have detections at $K$ or $K_{S}$. Eight such objects have $K/K_{S}$ nondetections establishing weak $K - $ch2 (or $K_S - $ch2) limits in the range of 1.97 to 3.14 magnitudes. A lack of bright $K/K_S$ counterparts is consistent with these eight $\geq$Y0 phototype targets being very late T or Y dwarfs.

Table \ref{tab:possible_y_dwarfs} lists 24 of our motion-confirmed discoveries which we believe merit further follow-up to better refute or confirm their status as Y dwarf candidates. There are three subclasses of objects listed in Table \ref{tab:possible_y_dwarfs}.  Within each of the three subclasses, our discoveries are listed from reddest (top) to bluest (bottom) in terms of ch1$-$ch2 color. The first 17 targets are those with \textit{Spitzer}-based phototypes $\geq$Y0. We include CWISEP 2355+3804 in this list even though its $J - $ch2 color suggests a more likely spectral type of late T. The second class of Y dwarf candidates listed are those sufficiently red to have ch1$-$ch2 colors within 1$\sigma$ of our phototyping grid's T/Y boundary at ch1$-$ch2 = 2.4 mag, excluding candidates with $J$ detections establishing $J - $ch2 $ < $ 5 mag. There are four such discoveries. The final class of possible Y dwarf candidates listed are those not within 1$\sigma$ of the T/Y color boundary, but nevertheless having measured ch1$-$ch2 redder than that of the bluest known Y dwarf in terms of this color\footnote{For the bluest ch1$-$ch2 color of any known Y dwarf, we adopt a value of ch1$-$ch2 = 2.11, corresponding to the Y0 dwarf WISE 2056+1459. WISE 1141$-$3326 (type Y0) has a bluer reported color of ch1$-$ch2 = $1.76 \pm 0.04$ \citep{davy_parallaxes}, but this anomaly is thought to result from blending with a background source \citep{tinney18}.} and lacking a $J$ detection establishing $J - $ch2 $ < $ 5 mag. Three of our discoveries fall in this category. We emphasize that Table \ref{tab:possible_y_dwarfs} targets in the latter two groups of Y dwarf candidates are likely to be late T dwarfs, but they nevertheless stand out as our most plausible Y dwarf candidates in the 2.1 $\lesssim$ ch1$-$ch2 $\lesssim$ 2.4 regime where T and Y dwarfs overlap considerably.

The Table \ref{tab:possible_y_dwarfs} column labeled [(ch1$-$ch2)$-$2.4]/$\sigma_{ch1-ch2}$ measures how much redder/bluer each candidate is than the adopted T/Y boundary at ch1$-$ch2 = 2.4, in units of each target's \textit{Spitzer} color uncertainty $\sigma_{ch1-ch2}$. This metric helps provide a sense for how certain we can be about whether each object is a T dwarf versus a Y dwarf. Five of our targets with phototype $\geq$Y0 are redder than our grid's T/Y boundary by more than 2$\sigma$. Much more follow-up is warranted in order to better measure/constrain the spectral types of the Y dwarf candidates presented in this section.

\subsection{Fastest-moving Discoveries}
\label{sec:fastest}

The median $\mu_{tot}$ of our motion-confirmed targets is 490 mas/yr. At the high motion tail of this distribution, three of our discoveries have best-fit $\mu_{tot} > 1.5''$/yr (CWISEP 0905$-$7400, CWISEP 1130+3139, and WISEA 1534$-$1043). An additional nine of our motion confirmed discoveries have $1.0''$/yr $ < \mu_{tot} < 1.5''$/yr. WISEA 1534$-$1043 has the highest total motion ($\mu_{tot} = 2697 \pm 68$ mas/yr) and reduced proper motion ($H_{ch2} = 22.92 \pm 0.06$ mag) of our motion-confirmed discoveries. The five largest $H_{ch2}$ values among our sample all belong to objects with ch1$-$ch$2 \le $ 1.26 mag, meaning that these targets occupy the bluest 20\% of our sample in terms of ch1$-$ch2 color. Figure \ref{fig:reduced_pm} shows a ch2 reduced proper motion diagram for all of our motion-confirmed targets with \textit{Spitzer} imaging available, plus the entire sample of Y dwarfs from the prior literature for context. While our targets with phototypes $\geq$Y0 generally inhabit the $18 \lesssim H_{ch2} \lesssim 21.5$ regime broadly consistent with that of literature Y dwarfs, none of our very red (ch1$-$ch2 $ > $ 2.4 mag) Y dwarf candidates comes within 2 mags of WISE 0855$-$0714 in terms of $H_{ch2}$.

Using the $\S$\ref{sec:abs_mags_dist} distance estimates, we obtain many $V_{tan}$ values well exceeding the typical `high kinematics' threshold of 100 km/s \citep[e.g.,][]{bdkp_part1}. 27 (13, 8, 4, 2) of our motion-confirmed discoveries have central $V_{tan}$ estimates in excess of 100 (150, 200, 300, 400) km/s. As discussed in $\S$\ref{sec:color_color}, we suspect that many of our highest $V_{tan}$ discoveries may be low-metallicity, subluminous objects, in which case their true tangential velocities would be lower than we have derived using the \cite{davy_parallaxes} $M_{ch2}$(ch1$-$ch2) relation.

\subsection{Nearest Discoveries}
\label{sec:nearest}

CWISEP 1935$-$1546 is our sample's only object with a central distance estimate placing it within 10 pc (Table \ref{tab:derived_properties}). However, with $d = 9.8^{+1.5}_{-1.3}$ pc, CWISEP 1935$-$1546 may well reside outside of the 10 pc volume. Meanwhile, the poorly constrained color of CWISEP 1446$-$2317 translates to a distance upper limit of $d < 8.3$ pc. Three additional discoveries have central distance estimates falling outside of 10 pc, but are within their 1$\sigma$ lower distance uncertainties of being closer than 10 pc (CWISEP 0402$-$2651, WISENF 1936+0408, CWISEP 2256+4002)

\subsection{The 20 pc Sample}

Completing the 20 pc census at types $\gtrsim$T5 represents a crucial step toward constraining late-type space densities, and ultimately the low-mass cutoff of the substellar mass function \citep[e.g.,][]{davy_parallaxes}. According to Table \ref{tab:derived_properties}, 21 of our motion-confirmed discoveries have central distance estimates (or a distance upper limit in the case of CWISEP 1446$-$2317) within 20 pc. These 21 targets have phototypes ranging from T7.5 to $\geq$Y1. An additional 24 of our motion-confirmed discoveries have central distance estimates larger than 20 pc, but are within their 1$\sigma$ lower distance uncertainties of being closer than 20 pc. 7 of our discoveries have 1$\sigma$ high distances still contained within 20 pc. Trigonometric distances will be needed to conclusively determine which of our discoveries indeed reside within 20 pc and take these new objects into account when computing space densities. In the absence of trigonometric distances, spectroscopic types could help refine absolute magnitude and hence distance estimates for our discoveries.

\subsection{NIR/Mid-IR Color-Color Plots}
\label{sec:color_color}

Our $\S$\ref{sec:ground_based} compilation of follow-up and archival $JHK/K_S$ photometry allows us to make a set of NIR/mid-IR color-color plots and thereby identify objects/populations with unusual spectral energy distributions. Figure \ref{fig:j_minus_ch2} provides a $J - $ch2 versus ch1$-$ch2 color-color diagram for motion-confirmed targets with \textit{Spitzer} photometry and $J$ band detections or limits available. By and large, our discoveries are in reasonable agreement with the literature trend for mid-late T dwarfs shown in magenta \citep{dupuy_liu_2012}. Our sample continues to display a rising $J - $ch2 trend toward the highest ch1$-$ch2 colors (latest spectral types), as anticipated. 

However, we note a distinct subpopulation with anomalously large $J - $ch2 colors ($J - $ch2 $ > $3.5 mag) given their relatively blue \textit{Spitzer} colors (ch1$-$ch2 $ < $ 1.5 mag) among our sample. Five of our discoveries are clearcut members of this color outlier population: CWISEP 0156+3255,  CWISEP 0505$-$5913, CWISEP 0700+7838, CWISEP 0905+7400 and WISEA 1534$-$1043. In all such cases, the measured $J - $ch2 colors/limits are at least 1.4 magnitudes redder than the \cite{dupuy_liu_2012} T dwarf trend (middle dashed magenta line in Figure \ref{fig:j_minus_ch2}).

There are several potential physical explanations for brown dwarf color outliers including metallicity, low/high gravity and binarity. Binarity is difficult to judge given our complete lack of trigonometric distances, spectroscopy and high-resolution imaging follow-up. On the other hand, kinematics can be helpful in assessing whether our color outliers may be unusually old (low metallicity, high gravity) or young (high metallicity, low gravity). Indeed, all five of our color outliers have unusually large reduced proper motions, $H_{ch2} > 21$ mag (see Figure \ref{fig:reduced_pm}). Only 10 of the 93 discoveries in Figure \ref{fig:j_minus_ch2} have $H_{ch2} > 21$ mag, and five of these are our color outliers, with another three much redder in ch1$-$ch2 phototyped as Y dwarfs. Additionally, all five color outliers have estimated $V_{tan} > 200$ km/s, suggesting they may be members of a halo population. In contrast, binarity would lead our $V_{tan}$ estimates to be biased low, and would also tend to favor relatively low $H_{ch2}$. Young objects would also preferentially have low kinematics. Thus, the high kinematics of our color outliers provide strong indications that this subsample is old, with relatively low metallicity and high gravity.

Our color outliers bear a striking resemblance to the benchmark T8 subdwarf WISE J200520.38+542433.9 \citep{wolf1130}, one of the best characterized mid-late T subdwarfs. Similar to our color outliers, WISE 2005+5424 is unusually red in $J - $ch2 ($J - $ch2 = $5.02 \pm 0.09$ mag) relative to its modest ch1$-$ch2 color of ch1$-$ch2 = $1.25 \pm 0.03$ mag. WISE 2005+5424 is shown in Figure \ref{fig:j_minus_ch2} as a yellow pentagon, and falls squarely amidst our five color outliers. As noted by \cite{wolf1130}, the models of \cite{burrows2006} do indeed predict that low metallicity objects with temperatures from 700-2200 K will have relatively red $J - $ch2 colors. We therefore believe that our color outliers are best considered mid-late T subdwarf candidates. If subdwarfs, their lower luminosities would correspond to smaller distances and hence help rein in their exceptionally large $V_{tan}$ values we have derived assuming an $M_{ch2}$(ch1$-$ch2) relation applicable to typical T dwarfs.

\cite{wolf1130} consider the ensemble of T dwarf color outliers, suggesting that $J - H$ color is useful in differentiating between old and young objects (see especially their Table 3 and Figure 8). Unfortunately, none of our five subdwarf candidates has an archival $H$ band detection available. Only WISEA 1534$-$1043 has an $H$ magnitude lower limit, and this corresponds to an uninformative color limit of $J - H < 1.98$ mag. It would be interesting to obtain deeper $H$ band follow-up of our five late-type subdwarf candidates, to determine whether they match the $J - H > -0.2$ mag trend noted by \cite{wolf1130} for old, low-metallicity T dwarfs.

Figure \ref{fig:hk_minus_ch2} shows color-color diagrams analogous to Figure \ref{fig:j_minus_ch2}, but for $H - $ch2 (left) and $K - $ch2 (right) rather than $J - $ch2. These plots provide at best a sanity check showing that the few targets with $H/K/K_S$ detections line up roughly along the expected trends for late type brown dwarfs. For the most part, our targets either have no archival $H/K/K_S$ imaging available or else obtain weak magnitude limits corresponding to essentially uninformative color limits.

\subsection{Notes on Individual Objects}

\subsubsection{CWISEP J021243.55+053147.2}
Among our 18 targets phototyped at $\geq$Y0, 17 have $\chi^2_{motion}$ values exceeding our significance threshold for motion confirmation. The remaining candidate with ch1$-$ch2 color in the Y dwarf regime is CWISEP 0212+0531. For this object, our \textit{Spitzer} astrometric data point clearly shows that despite its extremely red ch1$-$ch2 color, this object is consistent with being stationary (see Figure \ref{fig:radec_plots}). We do not count CWISEP 0212+0531 among our list of motion-confirmed Y dwarf candidates.

\subsubsection{CWISEP J022935.43+724616.4}
As noted in $\S$\ref{sec:spitzer_pairs}, CWISEP 0229+7246 has two ch2 counterparts but just one blended/elongated ch1 counterpart. We therefore consider it likely that CWISEP 0229+7246 is neither a single moving object nor a pair of closely spaced CPM objects, but rather some form of contaminant. For completeness, we can check the motion obtained by combining the ch2 center of light position with our W2 astrometry, as would be appropriate in the closely spaced CPM scenario. The resulting linear motion is $\mu_{\alpha} = 9 \pm 27$ mas/yr, $\mu_{\delta} = -29 \pm 26$ mas/yr. This corresponds to $\chi^2_{motion} = 1.4$, further disfavoring the close CPM system hypothesis.

\subsubsection{CWISEP J144606.62$-$231717.8}
\label{sec:j1446}

CWISEP 1446$-$2317 has the reddest best-fit ch1$-$ch2 color among our entire sample of brown dwarf candidates, with ch1$-$ch2 = $3.71 \pm 0.44$ mag. While this color nominally rivals that of WISE 0855$-$0714, the reddest known brown dwarf in ch1$-$ch2 and also the coldest known brown dwarf, we cannot claim that CWISEP 1446$-$2317 may be similarly cold (see Marocco et al., in prep., for additional follow-up/characterization of this source). Our Gemini $J$ band imaging with FLAMINGOS-2 confirms that CWISEP 1446$-$2317 is a Y dwarf by establishing a 5$\sigma$ limit of $J > 22.36$ mag, which corresponds to a 5$\sigma$ color lower limit of $J - $ch2 $ > 6.56$ mag.

Given our photometric distance constraint of $d < 8.3$ pc ($\S$\ref{sec:abs_mags_dist}), we sought to perform a custom WISE+\textit{Spitzer} astrometric analysis capable of enabling a combined parallax and proper motion measurement. Our typical linear motion fitting approach ($\S$\ref{sec:astrometry}) relies on extracting W2 detections from coadds spanning multiple WISE sky passes, posing challenges for parallax measurement, which is best suited to fitting WISE astrometry from single exposures or sky passes where the spacecraft position is well-defined. Although CWISEP 1446$-$2317 is extremely faint at W2 $ \sim 16$, we nevertheless performed astrometry only on W2 coadds of single WISE sky passes for this target. We also folded in the recently released 2018 NEOWISE W2 imaging for CWISEP 1446$-$2317 so as to leverage these two additional WISE sky passes. Using the \cite{ned_custom_astrometry} WISE coaddition and source detection methodology, we obtained W2 detections during 11 of the 12 available WISE sky passes throughout the 2010 to 2018 time period. These positional measurements are provided in Table \ref{tab:wise_positions}. We were unable to extract a CWISEP 1446$-$2317 W2 detection corresponding to the 2017 February NEOWISE sky pass.  

We combined our 11 epochs of WISE astrometry with our \textit{Spitzer} ch2 position from Table \ref{tab:spitzer_positions} to perform a parallax plus proper motion fit accounting for the WISE and \textit{Spitzer} ephemerides. These yielded a motion solution with $\mu_{\alpha} = -774 \pm 82$ mas/yr, $\mu_{\delta} = -978 \pm 71$ mas/yr and a parallax of $-501 \pm 248$ mas, with $\chi^2 = 17.4$ for 19 dof. The negative parallax is clearly neither physical nor statistically significant. Consequently, we revert to the simple linear motion fitting used for the rest of our sample ($\S$\ref{sec:poly_fit}) when quoting the CWISEP 1446$-$2317 motion in Table \ref{tab:wise_spitzer_pm}. Regardless of the fitting details, the CWISEP 1446$-$2317 linear motion is detected at very high significance.

We also checked the CatWISE \verb|par_pm| parameter for CWISEP 1446$-$2317. \verb|par_pm| provides a parallax estimate based only on WISE data spanning 2010-2016 by comparing linear motion fits computed separately for ascending and descending scans, with WISE observing from $\sim$1 AU on opposite sides of the Sun during opposite scan directions. CWISEP 1446$-$2317 has CatWISE \verb|par_pm| = $253 \pm 329$ mas, which is again not statistically significant. More \textit{Spitzer}-precision astrometric follow-up will be needed to obtain a reliable trigonometric parallax for this source (see Marocco et al., in prep., for additional CWISEP 1446$-$2317 astrometry).

\begin{figure*}
\plotone{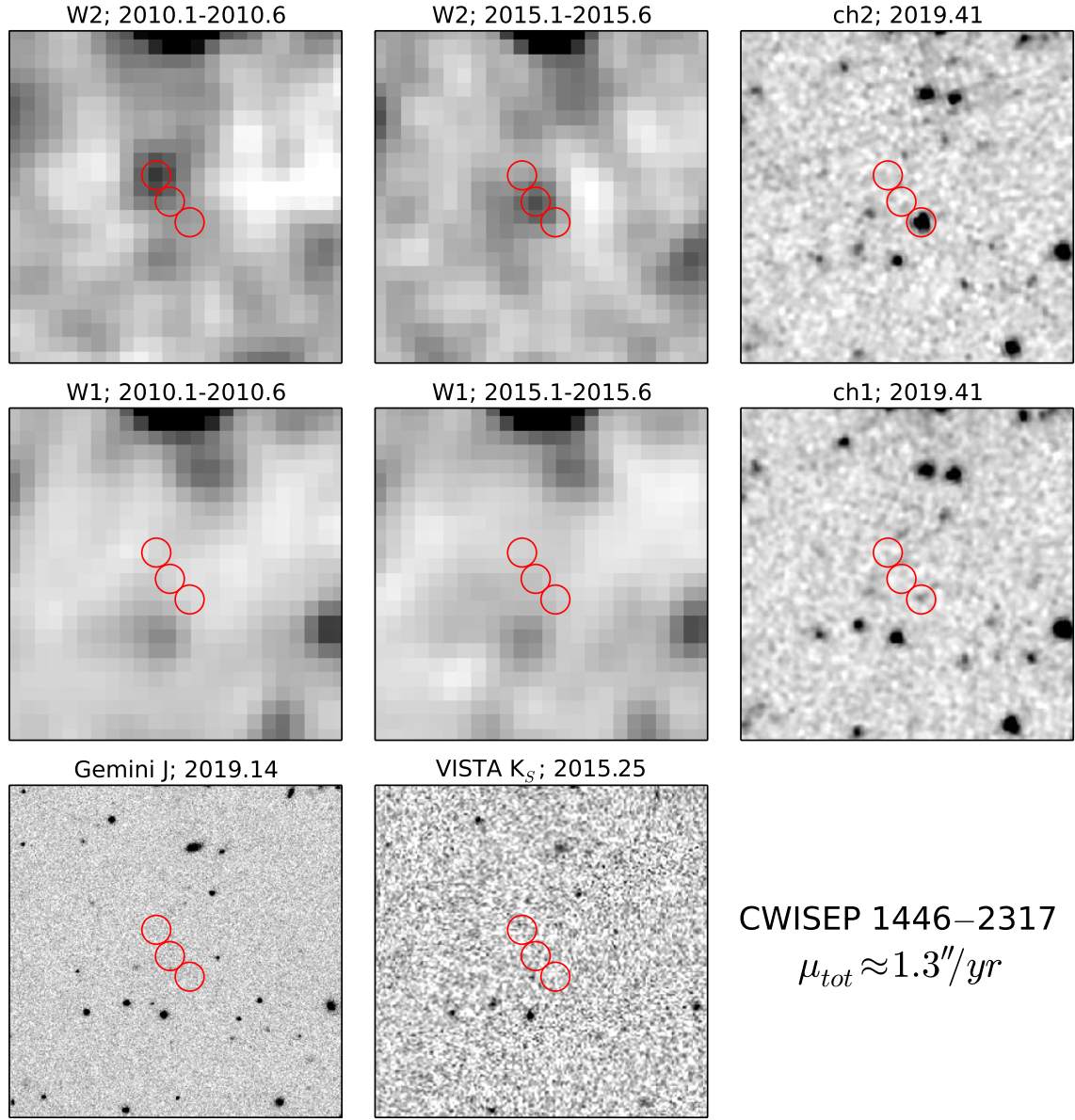}
\caption{Postage stamps showing available archival and follow-up imaging for CWISEP 1446$-$2317, our sample's reddest target in terms of best-fit ch1$-$ch2 color. Each cutout is 1.1$'$ on a side. East is left and north is up. Within each panel, the red circles indicate the 2010.1-2010.6 WISE position (northeastern circle), the 2015.1-2015.6 WISE position (central circle) and the 2019.41 \textit{Spitzer} position (southwestern circle). CWISEP 1446$-$2317 appears undetected in W1, Gemini/FLAMINGOS-2 $J$ band follow-up (PI Gelino), and archival VISTA $K_S$ imaging from VHS. \textit{Spitzer} imaging comes from our p14034 campaign. The ch1 detection is very weak, leading to a very large uncertainty on the measured ch1$-$ch2 color. The WISE images are one-year unWISE meta-coadds, spanning calendar 2010 (2015) in the leftmost (center) column, and have been smoothed by a $6.5''$ FWHM Gaussian kernel.}
\label{fig:j1446_cutouts}
\end{figure*}

Figure \ref{fig:j1446_cutouts} shows WISE, \textit{Spitzer}, Gemini and VISTA image cutouts illustrating the motion and colors of CWISEP 1446$-$2317. Note that this figure only shows two years out of the six total years worth of W1/W2 imaging available. The three red circles remain fixed in all panels at the 2010 WISE, 2015 WISE and 2019.41 \textit{Spitzer} positions. There is no 2019.41 \textit{Spitzer} counterpart at the 2010 WISE position, nor a WISE counterpart during either 2011 or 2015 at the \textit{Spitzer} 2019.41 position. The 2015 W2 position does not appear perfectly consistent with a linear interpolation between the 2010 W2 and 2019.41 \textit{Spitzer} positions, seemingly dragged slightly southeast by a noise excursion. This underscores the difficulty of performing astrometry for such a faint W2 source so close to the background noise limit. There is no trace of a counterpart in either W1, Gemini $J$ or VISTA $K_{S}$. The ch1 counterpart shown is very weak, leading to the large uncertainty on our measured ch1$-$ch2 color. VISTA provides only a weak $K_S - $ch2 $ > $ 2.06 mag limit.

Lastly, it is interesting to compare the reduced proper motion of CWISEP 1446$-$2317 ($H_{ch2} = 21.35 \pm 0.12$ mag) to that of the previously known Y dwarf sample. Among all Y dwarfs from the prior literature, only WISE 0855$-$0714 has a larger $H_{ch2}$ than does CWISEP 1446$-$2317, as can be seen in Figure \ref{fig:reduced_pm}. However, CWISEP 1446$-$2317 is a very distant second, with the WISE 0855$-$0714 $H_{ch2}$ being more than 2 magnitudes larger.

\subsubsection{WISEA J153429.75$-$104303.3}
\label{sec:the_accident}

WISEA 1534$-$1043 is the fastest-moving discovery in our sample by multiple metrics: total linear motion ($\mu_{tot} = 2697 \pm 68$ mas/yr), ch2 reduced proper motion ($H_{ch2} = 22.92 \pm 0.06$ mag) and estimated tangential velocity ($V_{tan} = 485^{+73}_{-64}$ km/s). With $J - $ch2 $ > 4.79$ mag and ch1$-$ch2 = $0.93 \pm 0.04$  mag, WISEA 1534$-$1043 joins a small subpopulation of color outliers within our motion-confirmed sample as illustrated in Figure \ref{fig:j_minus_ch2}. As discussed in $\S$\ref{sec:color_color}, we favor an explanation for the unusual properties of WISEA 1534$-$1043 in which it is a mid-late T subdwarf, based on kinematics and the striking similarity of our color outlier population to the benchmark T8 subdwarf WISE 2005+5424. WISEA 1534$-$1043 is entirely undetected in our Palomar/WIRC $J$ band follow-up, and deeper $J$ band imaging would be useful toward establishing just how anomalously red this object is in its $J - $ch2 color. If WISEA 1534$-$1043 is indeed a subdwarf, that would place it at a closer distance than our current $38.0^{+5.6}_{-4.9}$ pc estimate, and correspondingly reduce its exceptionally large $V_{tan}$ estimate.

Figure \ref{fig:radec_plots} includes a scatter plot displaying the measured (RA, Dec) trajectory of WISEA 1534$-$1043. For WISEA 1534$-$1043, we attempted the same WISE+\textit{Spitzer} parallax fitting methodology
as used in $\S$\ref{sec:j1446} for CWISEP 1446$-$2317. In the case of WISEA 1534$-$1043, the joint parallax plus proper motion fit yields $\mu_{\alpha} = -1254 \pm 116$ mas/yr, $\mu_{\delta} = -2406 \pm 84$ mas/yr and parallax of $329 \pm 422$ mas, with $\chi^2 = 8.5$ for 9 dof. The lower number of dof results from the fact that we could only extract single sky pass W2 detections for 7 of 12 WISE sky passes. Our parallax measurement is not at all statistically significant. No CatWISE \verb|par_pm| parallax estimate is available for WISEA 1534$-$1043, since this target is wholly absent from the CatWISE catalog. More high-precision astrometric follow-up of WISEA 1534$-$1043 would help pin down its ch2 absolute magnitude and thereby test our subdwarf hypothesis.

\begin{figure*}
\plotone{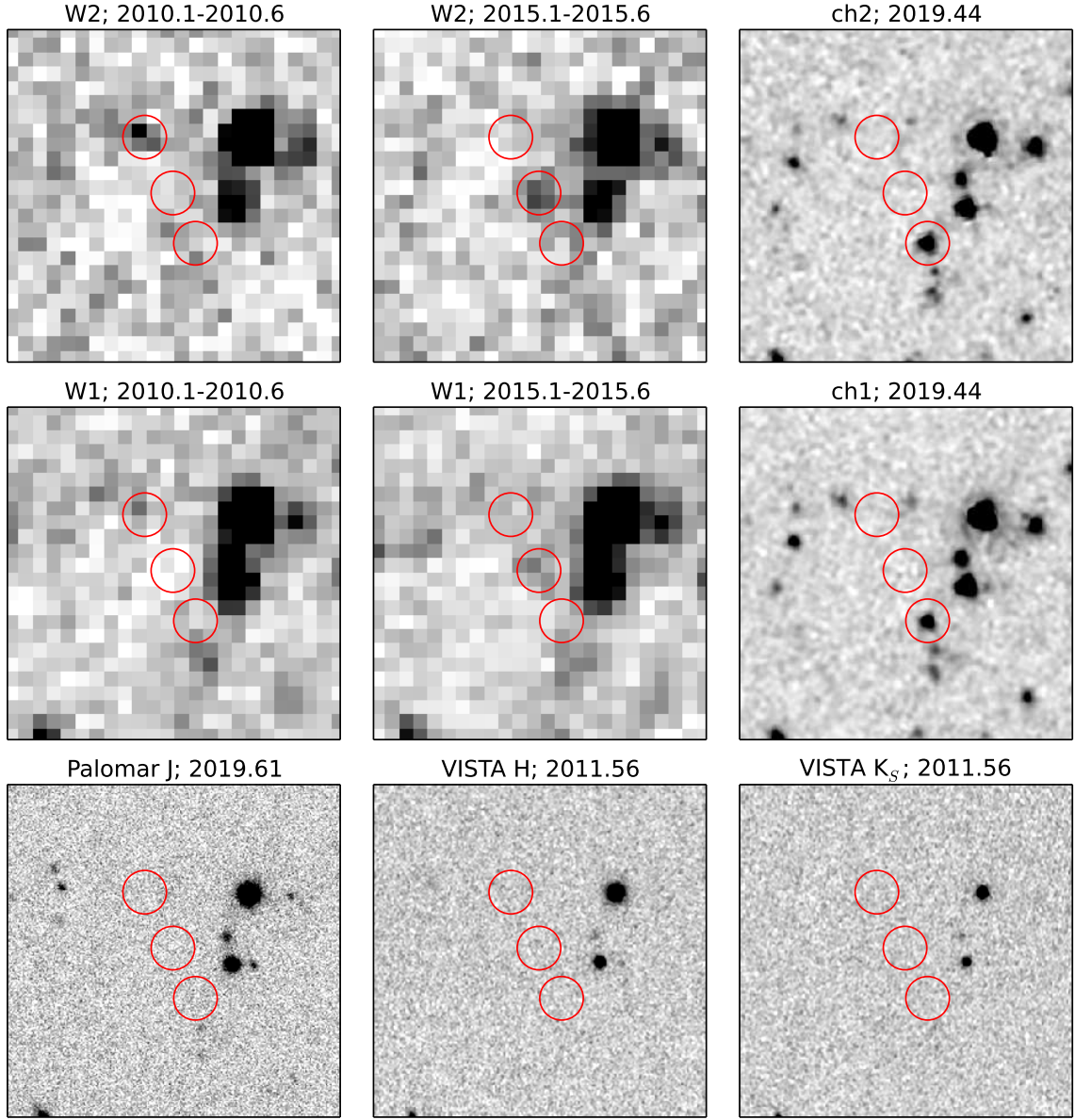}
\caption{Postage stamps showing available archival and follow-up imaging for WISEA 1534$-$1043, our sample's highest proper motion target ($\mu_{tot} \approx 2.7''$/yr), which also has the largest ch2 reduced proper motion of any target in our sample ($H_{ch2} \approx 22.9$ mag). Each cutout is 1.1$'$ on a side. East is left and north is up. Within each panel, the red circles indicate the 2010.1-2010.6 WISE position (northeastern circle), the 2015.1-2015.6 WISE position (central circle) and the 2019.44 \textit{Spitzer} position (southwestern circle). WISEA 1534$-$1043 has an extremely faint W1 counterpart. It is undetected in our Palomar/WIRC $J$ band follow-up (PI Marocco) and archival VISTA $H$, $K_S$ imaging from VHS. \textit{Spitzer} imaging comes from our p14034 campaign. Interestingly, this source has a relatively bright ch1 counterpart, occupying an unusual region of color-color space with ch1$-$ch2 $ < 1$ and $J - $ch2 $ > 4.5$. The WISE images are one-year unWISE meta-coadds, spanning calendar 2010 (2015) in the leftmost (center) column. In this case the WISE images are kept unsmoothed for clarity because the target moves into a position adjacent to much brighter static sources over time.}
\label{fig:j1534_cutouts}
\end{figure*}

Figure \ref{fig:j1534_cutouts} shows WISE, \textit{Spitzer}, Palomar, and VISTA image cutouts in the vicinity of WISEA 1534$-$1043. In this case, we do not smooth the WISE images since the moving target shifts into a position adjacent to much brighter static sources over time. WISEA 1534$-$1043 has an extremely faint, but nonetheless legitimate, W1 counterpart. WISEA 1534$-$1043 is quite blue in ch1$-$ch2 relative to the bulk of our sample, and it is very well detected in both \textit{Spitzer} channels. There is no trace of a counterpart detection in the NIR panels at bottom. As mentioned in $\S$\ref{sec:color_color}, deeper $H$ band follow-up would be of interest in determining whether WISEA 1534$-$1043 matches the $J - H > -0.2 $ mag trend noted by \cite{wolf1130} for mid-late T subdwarfs.

\subsubsection{CWISEP J154151.59+523025.0}

This target has two \textit{Spitzer} counterparts with similar magnitudes in both ch1 and ch2. Further complicating matters, the potentially moving W2 source appears contaminated by blending with a much bluer stationary object. Additional follow-up would be needed to conclusively determine whether CWISEP 1541+5230 is a pair of red extragalactic sources or instead two companion brown dwarfs. Both \textit{Spitzer} counterparts, if assumed to be late-type brown dwarfs, have photometric distances consistent with $d \approx$  38 pc and T8 spectral type estimates. The two \textit{Spitzer} sources are separated by $3.65''$, which would translate to a projected physical separation of 139 AU using $d = 38$ pc. If we assume that CWISEP 1541+5230 is a CPM system, then we obtain a linear motion of $\mu_{\alpha} = 208 \pm 27$ mas/yr, $\mu_{\delta} = -209 \pm 29$ mas/yr by fitting the W2 astrometry in combination with the ch2 center of light position.

\subsubsection{CWISEP J193518.59$-$154620.3}
Our p14034 photometry of CWISEP 1935$-$1546 yielded an exceptionally red, though very noisy, \textit{Spitzer} color estimate of $3.24 \pm 0.31$ mag \citep{marocco2019}. Because this color interval overlapped with that of the coldest known brown dwarf (WISE 0855$-$0714; ch1$-$ch2 = $3.55\pm 0.07$), we obtained much deeper ch1 imaging via \textit{Spitzer} DDT program 14279 (p14279; PI Marocco). We also obtained additional ch2 astrometry via observations executed immediately `back-to-back' with those of p14279, allowing us to measure a much higher S/N color based on nearly simultaneous imaging in both IRAC bands. These additional ch2 astrometric observations were part of DDT program 14224 (p14224; PI Kirkpatrick). Our total p14279 ch1 integration time was 3600 seconds, as compared to just 210 seconds with p14034.

Combining the p14279 and p14224 data acquired on 2019 August 7, we obtain a refined color measurement of ch1$-$ch2 = $2.984 \pm 0.034$ mag. So the original CWISEP 1935$-$1546 color from p14034 appears to have scattered $\sim$1$\sigma$ red relative to the more accurate value now in hand. In Table \ref{tab:wise_spitzer_photom} we list the newer, higher S/N CWISEP 1935$-$1546 ch1$-$ch2 color rather than that from p14034. This color places CWISEP 1935$-$1546 squarely in the Y1 phototype regime, with a corresponding photometric distance estimate of 9.8$^{+1.5}_{-1.3}$ pc. This also corresponds to a $V_{tan} = 14^{+4}_{-3}$ km/s estimate significantly refined relative to that initially presented in \cite{marocco2019}. Despite being the third reddest motion-confirmed discovery in our entire sample by central ch1$-$ch2 color, CWISEP 1935$-$1546 has the lowest $H_{ch2}$ among our motion-confirmed targets with $\geq$Y0 phototypes.

\section{Conclusion}
\label{sec:conclusion}
We have undertaken an extensive effort to mine the CatWISE proper motion catalog, and more generally the combined WISE+NEOWISE data set, to discover extremely cold brown dwarfs hitherto overlooked by prior searches. By leveraging the 6.5+ year time baseline afforded by the combination of pre-hibernation and post-reactivation W1/W2 imaging, we are able to perform much deeper motion-based brown dwarf selections than prior all-sky WISE moving object surveys. As a result, we have discovered many faint moving objects with exceptionally red \textit{Spitzer} colors and hence extremely cold temperatures. Based on \textit{Spitzer} ch1$-$ch2 color measurements and WISE+\textit{Spitzer} motion validation, our sample contains 17 newly discovered motion-confirmed brown dwarfs with best-fit spectral types $\ge$Y0, 16 of which have $J - $ch2 detections/limits consistent with being Y dwarfs. Much more follow-up will be required in order to determine which of our Y dwarf candidates listed in Table \ref{tab:possible_y_dwarfs} are indeed Y dwarfs, and ultimately obtain spectral types.

One of our Y dwarf candidates, CWISEP 1446$-$2317, stands out from the rest with an exceptionally large though very uncertain \textit{Spitzer} color of ch1$-$ch2 = $3.71 \pm 0.44$. This source in particular merits additional follow-up to refine its photometry and astrometry (see Marocco et al., in prep., for additional characterization of this source).

Table \ref{tab:possible_y_dwarfs} illustrates that deeper $J$ band follow-up will help push many of our Y dwarf candidates securely into a $J - $ch2 regime so red as to exclude late T dwarfs. Further astrometric follow-up will be highly valuable on a number of fronts. Trigonometric parallaxes would provide absolute magnitudes which can discriminate between late T and Y spectral types. Trigonometric parallaxes will also be necessary to determine which of our discoveries fall within the 20 pc volume and incorporate these into accurate space density estimates. Deeper $H$ band imaging follow-up would be particularly useful in determining whether our sample of fast-moving color outliers are indeed mid-late T subdwarfs as we have hypothesized.

We anticipate many more very late type CatWISE brown dwarf discoveries in the future. A second and final CatWISE data processing is underway, expected to catalog an additional $\sim$1 billion WISE-selected objects by virtue of more aggressive source detection/deblending. However, \textit{Spitzer} is a uniquely efficient resource for obtaining spectral type estimates of such discoveries, and upon \textit{Spitzer}'s retirement it will become much more challenging to sift for promising Y dwarf candidates among W2-only moving objects.

The upcoming decade will provide exciting opportunities to discover and characterize the coldest ($T_{eff} \lesssim 300$ K) substellar constituents of the solar neighborhood \citep[e.g.,][]{davy_white_paper, leggett_white_paper}. The community will need to select JWST Y dwarf targets of maximum scientific value, to better understand both brown dwarfs themselves and giant exoplanet atmospheres \citep{leggett_white_paper}. The Near Earth Object Surveyor (formerly NEOCam) will be able to reveal colder, fainter, and more distant Y dwarfs than WISE given resources for coadded processings analogous to those of AllWISE/unWISE/CatWISE \citep{davy_white_paper}. Determining the low mass cutoff of star formation will require a complete sample out to distances of 20-50 pc \citep{leggett_white_paper}, far beyond the volume currently mapped \citep{davy_parallaxes}. CatWISE is already making progress on these fronts by mining the WISE/NEOWISE data set for Y dwarf discoveries to its faintest attainable depths.

\acknowledgments

We wish to thank the anonymous referee. We thank Nicholas Cross for assistance with compiling archival UKIRT/VISTA detections. This research was partially carried out at the Jet Propulsion Laboratory, California Institute of Technology, under a contract with NASA.

CatWISE is funded by NASA under Proposal No. 16-ADAP16-0077 issued through the Astrophysics Data Analysis Program, and uses data from the NASA-funded WISE and NEOWISE projects. AMM acknowledges support from Hubble Fellowship HST-HF2-51415.001-A. FM is supported by an appointment to the NASA Postdoctoral Program at the Jet Propulsion Laboratory, administered by Universities Space Research Association under contract with NASA.

This work is based in part on observations made with the \textit{Spitzer} Space Telescope, which is operated by the Jet Propulsion Laboratory, California Institute of Technology under a contract with NASA.

Based in part on observations obtained at the Gemini Observatory, which is operated by the Association of Universities for Research in Astronomy, Inc., under a cooperative agreement with the NSF on behalf of the Gemini partnership: the National Science Foundation (United States), National Research Council (Canada), CONICYT (Chile), Ministerio de Ciencia, Tecnolog\'{i}a e Innovaci\'{o}n Productiva (Argentina), Minist\'{e}rio da Ci\^{e}ncia, Tecnologia e Inova\c{c}\~{a}o (Brazil), and Korea Astronomy and Space Science Institute (Republic of Korea).

\vspace{5mm}
\facilities{Spitzer(IRAC), WISE/NEOWISE, \\ Gemini(FLAMINGOS-2),
Hale(WIRC), UKIRT(WFCAM), VISTA(VIRCAM), 2MASS}

\software{XGBoost \citep{xgboost}, MOPEX \citep{mopex, mopex_extraction}, WiseView \citep{wiseview}}

\bibliography{catwise_p14034}

\movetabledown=13mm
\begin{longrotatetable}
\begin{deluxetable*}{lccccccccc}
\tablecaption{WISE \& \textit{Spitzer} Photometry \label{tab:wise_spitzer_photom}}
\tablehead{
\colhead{Name} & \colhead{motion} & \colhead{W1} & \colhead{W2} & \colhead{W1$-$W2} & \colhead{AOR} & \colhead{ch1} & \colhead{ch2} & \colhead{ch1$-$ch2} & \colhead{search}  \\
 & \colhead{confirmed} & \colhead{(mag)} & \colhead{(mag)}  & \colhead{(mag)} & & \colhead{(mag)} & \colhead{(mag)} & \colhead{(mag)} & \colhead{method\tablenotemark{c}} }
\startdata
CWISEP J000006.01$-$704851.2 & yes & $18.390 \pm 0.153$ & $15.834 \pm 0.051$ & $2.556 \pm 0.161$ & 68730112 & $17.293 \pm 0.058$ & $15.836 \pm 0.022$ & $1.457 \pm 0.062$ & M2 \\
CWISEP J000110.81$-$093215.5 & yes & $18.598 \pm 0.259$ & $16.104 \pm 0.086$ & $2.494 \pm 0.273$ & 68727808 & $17.573 \pm 0.065$ & $15.889 \pm 0.021$ & $1.684 \pm 0.068$ & M2 \\
CWISEP J001146.07$-$471306.8 & yes & $18.939 \pm 0.272$ & $15.986 \pm 0.062$ & $2.953 \pm 0.279$ & 68486656 & $17.744 \pm 0.090$ & $15.812 \pm 0.021$ & $1.931 \pm 0.093$ & M1 \\
CWISEP J003507.77$-$153233.8 & yes & $17.351 \pm 0.080$\tablenotemark{a} & $15.252 \pm 0.039$ & $2.099 \pm 0.089$ & 68235776 & $17.111 \pm 0.065$ & $15.228 \pm 0.020$ & $1.884 \pm 0.068$ & C6 \\
CWISEP J003915.43+360939.0 & no & $>19.043$ & $16.303 \pm 0.087$ & $>2.740$ & 68506112 & $18.191 \pm 0.102$ & $15.982 \pm 0.021$ & $2.210 \pm 0.104$ & C8 \\
CWISEP J004158.35+381811.9 & yes & $18.654 \pm 0.248$ & $15.957 \pm 0.065$ & $2.697 \pm 0.256$ & 68486144 & $17.817 \pm 0.095$ & $16.069 \pm 0.024$ & $1.749 \pm 0.098$ & M1 \\
CWISEP J005802.63+723330.3 & yes & $17.395 \pm 0.072$ & $16.109 \pm 0.065$ & $1.286 \pm 0.097$ & 68731904 & $17.500 \pm 0.062$ & $16.749 \pm 0.034$ & $0.751 \pm 0.070$ & C3 \\
CWISEP J010247.48$-$654226.4 & no & $19.416 \pm 0.512$ & $16.681 \pm 0.099$ & $2.735 \pm 0.521$ & 68509952 & $18.787 \pm 0.159$ & $16.432 \pm 0.023$ & $2.354 \pm 0.160$ & C8 \\
\enddata
\tablenotetext{a}{WISE photometry may be contaminated by blending.}
\tablenotetext{b}{Object is scheduled for p14034 \textit{Spitzer} observations in the near future, but does not yet have \textit{Spitzer} photometry available.}
\tablenotetext{c}{The search method names listed in this column are defined in $\S$\ref{sec:catwise_searches}-\ref{sec:unwise_searches}, which also contain explanations of each search's moving object candidate selection procedure.}
\end{deluxetable*}
\end{longrotatetable}

\begin{deluxetable}{cll}
\tablecaption{Spurious Candidates \label{tab:duds}}
\tablehead{
\colhead{p14034 AOR} & \colhead{CWISEP designation} & \colhead{W2} \\
\colhead{} & \colhead{} & \colhead{(mag)}
}
\startdata
68510720 & J075426.59$-$025947.8 & $16.36 \pm 0.12$ \\
68502784 & J111337.00+342927.0 & $16.48 \pm 0.13$ \\
68505600 & J121005.73$-$150533.3 & $15.97 \pm 0.08$ \\
68733696 & J132904.57+343524.4 & $16.31 \pm 0.08$ \\
68577024 & J202736.23$-$621855.5 & $16.68 \pm 0.14$ \\
68506880 & J212111.15+005640.0 & $16.34 \pm 0.10$ \\
\enddata
\end{deluxetable}

\startlongtable
\begin{deluxetable*}{lcccccc}
\tablecaption{WISE W2 Positions used in Joint WISE+\textit{Spitzer} Linear Motion Fits \label{tab:wise_positions}}
\tablehead{
\colhead{Name} & \colhead{RA} & \colhead{Dec} & \colhead{$\sigma_{RA}$} & \colhead{$\sigma_{Dec}$} & \colhead{time} & \colhead{mean}  \\
 & \colhead{(deg, ICRS)} & \colhead{(deg, ICRS)} & \colhead{(mas)} & \colhead{(mas)} & \colhead{slice\tablenotemark{a}} & \colhead{MJD} 
}
\startdata
CWISEP J000006.01$-$704851.2 & 0.025954 & $-$70.814036 & 424 & 408 & pre       & 55404.23 \\
                           & 0.024768 & $-$70.814293 & 283 & 267 & post      & 57330.72 \\
CWISEP J000110.81$-$093215.5 & 0.294481 & $-$9.537603 & 678 & 691 & pre       & 55431.64 \\
                           & 0.295332 & $-$9.537668 & 539 & 540 & post      & 57387.92 \\
CWISEP J001146.07$-$471306.8 & 2.941300 & $-$47.218389 & 620 & 625 & pre       & 55442.17 \\
                           & 2.942075 & $-$47.218477 & 364 & 361 & post      & 57398.73 \\
CWISEP J003507.77$-$153233.8 & 8.782612 & $-$15.542104 & 422 & 426 & pre       & 55447.53 \\
                           & 8.782597 & $-$15.542193 & 481 & 469 & post0\_1yr & 56747.59 \\
\enddata
\tablenotetext{a}{See $\S$\ref{sec:wise_astrometry} and Table \ref{tab:slices} for discussion and definitions of the WISE time slices used for W2 astrometry.}
\end{deluxetable*}

\startlongtable
\begin{deluxetable*}{lcccccccc}
\tablecaption{Recalibrated \textit{Spitzer} ch2 Positions \label{tab:spitzer_positions}}
\tablehead{
\colhead{Name} & \colhead{AOR} & \colhead{RA} & \colhead{Dec} & \colhead{$\sigma_{RA}$} & \colhead{$\sigma_{Dec}$} & \colhead{MJD} & \colhead{method} & \colhead{$N_{calib}$} \\
 & & \colhead{(deg, ICRS)} & \colhead{(deg, ICRS)} & \colhead{(mas)} & \colhead{(mas)} &  & \colhead{number\tablenotemark{a}} & 
}
\startdata
CWISEP J000006.01$-$704851.2   & 68730112 & 0.02385537 & $-$70.81443752 & 42 & 36 & 58711.09 &        2 &           13 \\
CWISEP J000110.81$-$093215.5   & 68727808 & 0.29543643 & $-$9.53797990 & 25 & 27 & 58559.84 &        3 &            8 \\
CWISEP J001146.07$-$471306.8   & 68486656 & 2.94245662 & $-$47.21861888 & 47 & 59 & 58531.38 &        3 &            9 \\
CWISEP J003507.77$-$153233.8   & 68235776 & 8.78254804 & $-$15.54325614 & 25 & 32 & 58427.81 &        3 &            6 \\
CWISEP J003915.43+360939.0   & 68506112 & 9.81402661 & 36.16052959 & 26 & 23 & 58583.45 &        1 &           12 \\
CWISEP J004158.35+381811.9   & 68486144 & 10.49371897 & 38.30336814 & 35 & 42 & 58468.55 &        1 &           10 \\
CWISEP J005802.63+723330.3   & 68731904 & 14.51001015 & 72.55856213 & 65 & 54 & 58594.58 &        1 &           35 \\
CWISEP J010247.48$-$654226.4   & 68509952 & 15.69697472 & $-$65.70750484 & 62 & 40 & 58473.60 &        6 &           12 \\
\enddata
\tablecomments{$N_{calib}$ is the number of \textit{Gaia} DR2 calibrators used for \textit{Spitzer} ch2 astrometric recalibration of each AOR, as discussed in $\S$\ref{sec:spitzer_astrom_recalib}.}
\tablenotetext{a}{The astrometric calibrator selection `method number' here refers to the method number listed in Table \ref{tab:spitzer_recalib_samples} and explained in $\S$\ref{sec:spitzer_astrom_recalib}.}
\end{deluxetable*}

\begin{deluxetable}{ll}
\tablecaption{unWISE Meta-coadd Time Slices \label{tab:slices}}
\tablehead{
\colhead{Time Slice Name} & \colhead{approx. time period} \\
\colhead{} & \colhead{(calendar years)}
}
\startdata
pre & 2010.0-2011.1 \\
post & 2014 \& 2015 \& 2016 \& 2017 \\
post0\_1yr & 2014 \\
post1\_1yr & 2015 \\
post2\_1yr & 2016 \\
post3\_1yr & 2017 \\
post0\_2yr & 2014 \& 2015 \\
post1\_2yr & 2016 \& 2017 \\
\enddata
\end{deluxetable}

\begin{deluxetable*}{cccccc}
\tablecaption{Cuts Defining \textit{Spitzer}-\textit{Gaia} Astrometric Calibrator Samples \label{tab:spitzer_recalib_samples}}
\tablehead{
\colhead{method number} & \colhead{min. ch2 SNR} & \colhead{min. BCD cov. frac.} & \colhead{\textit{Gaia} PM required} & \colhead{$N_{calib, min}$} & \colhead{number of AORs}
}
\startdata
       1 & 100 & 1.0 & yes &       10&           67 \\
       2 & 50 & 1.0 & yes &       10&           28 \\
       3 & 30 & 1.0 & yes &        5&           57 \\ \hline
       4 & 100 & 1.0 & no &       10&            1 \\
       5 & 50 & 1.0 & no &       10&            0 \\
       6 & 30 & 1.0 & no &        5&            6 \\ \hline
       7 & 100 & 0.5 & yes &       10&            0 \\
       8 & 50 & 0.5 & yes &       10&            0 \\
       9 & 30 & 0.5 & yes &        5&            5 \\
\enddata
\end{deluxetable*}

\movetabledown=7mm
\begin{longrotatetable}
\begin{deluxetable*}{lcccccccccccc}
\tablecaption{WISE+\textit{Spitzer} Linear Motion Fitting Results \label{tab:wise_spitzer_pm}}
\tablehead{
\colhead{Name} & \colhead{$\alpha_0$} & \colhead{$\delta_0$} & \colhead{$\sigma_{\alpha_0}$} & \colhead{$\sigma_{\delta_0}$} & \colhead{MJD$_0$} & \colhead{$\mu_{\alpha}$} & \colhead{$\mu_{\delta}$} & \colhead{$\mu_{tot}$} & \colhead{$\chi^2_{motion}$} & \colhead{$\chi^2$} & \colhead{\# dof} & \colhead{$\chi^2_{\nu}$} \\
 & \colhead{(deg, ICRS)} & \colhead{(deg, ICRS)} & \colhead{(mas)}  & \colhead{(mas)} & & \colhead{mas/yr} & \colhead{mas/yr} & \colhead{mas/yr} & & & & }
\startdata
CWISEP J000006.01$-$704851.2 & 0.0238910 & $-$70.8144313 & 42 & 36 & 58656.07 & $-277 \pm 40$ & $-153 \pm 38$ & $317 \pm 40$ & 63.3 & 0.08 & 2 & 0.04 \\
CWISEP J000110.81$-$093215.5 & 0.2954347 & $-$9.5379786 & 26 & 27 & 58552.36 & $345 \pm 72$ & $-194 \pm 73$ & $396 \pm 72$ & 30.2 & 3.34 & 2 & 1.67 \\
CWISEP J001146.07$-$471306.8 & 2.9424406 & $-$47.2186142 & 47 & 58 & 58486.49 & $325 \pm 63$ & $-117 \pm 63$ & $346 \pm 63$ & 30.2 & 0.29 & 2 & 0.15 \\
CWISEP J003507.77$-$153233.8 & 8.7825466 & $-$15.5432421 & 25 & 32 & 58394.44 & $20 \pm 41$ & $-553 \pm 41$ & $553 \pm 41$ & 181.8 & 14.45 & 8 & 1.81 \\
CWISEP J003915.43+360939.0 & 9.8140272 & 36.1605309 & 27 & 23 & 58577.21 & $-127 \pm 78$ & $-227 \pm 79$ & $260 \pm 79$ & 10.9 & 1.75 & 2 & 0.87 \\
CWISEP J004158.35+381811.9 & 10.4937097 & 38.3033679 & 35 & 42 & 58437.39 & $297 \pm 51$ & $-8 \pm 52$ & $297 \pm 51$ & 34.1 & 0.54 & 2 & 0.27 \\
CWISEP J005802.63+723330.3 & 14.5100692 & 72.5585540 & 64 & 54 & 58534.48 & $-474 \pm 61$ & $196 \pm 59$ & $513 \pm 60$ & 72.1 & 6.72 & 2 & 3.36 \\
CWISEP J010247.48$-$654226.4 & 15.6969790 & $-$65.7075044 & 62 & 41 & 58453.38 & $-186 \pm 85$ & $-35 \pm 82$ & $189 \pm 85$ & 5.0 & 1.88 & 6 & 0.31 \\
\enddata
\tablecomments{$\chi^2_{\nu}$ is the reduced $\chi^2$ of the linear motion fit, defined as $\chi^2_{\nu} = \chi^2/$(\# dof).}
\end{deluxetable*}
\end{longrotatetable}

\begin{deluxetable*}{cccccc}
\tablecaption{2MASS Counterparts \label{tab:tmass}}
\tablehead{
\colhead{Name} & \colhead{2MASS} & \colhead{$J_{2MASS}$} & \colhead{$H_{2MASS}$} & \colhead{$K_{S, 2MASS}$}  \\
 & \colhead{desig.} & &  &   
}
\startdata
CWISEP J140247.83+102132.6 & 2MASS J14024796+1021318 & 16.443 $\pm$ 0.139 & $>$15.692 & $>$16.369 \\
CWISEP J201510.68$-$675005.6 & 2MASS J20151069$-$6750029 & 17.027 $\pm$ 0.181 & $>$16.731 & $>$16.374 \\
\enddata
\end{deluxetable*}

\movetabledown=7mm
\begin{longrotatetable}
\begin{deluxetable*}{lcccccccc}
\tablecaption{NIR photometry from Gemini, Palomar, UKIRT and VISTA for motion-confirmed targets \label{tab:jhk_phot}}
\tablehead{
\colhead{Name} & \colhead{$J_{MKO}$} & \colhead{origin} & \colhead{$H_{MKO}$} & \colhead{origin} & \colhead{$K_{MKO}$} & \colhead{origin} & \colhead{$K_S$} & \colhead{origin} \\
 & \colhead{(mag)} & \colhead{of $J_{MKO}$} & \colhead{(mag)}  & \colhead{of $H_{MKO}$} & \colhead{(mag)} & \colhead{of $K_{MKO}$} & \colhead{(mag)} & \colhead{of $K_S$} }
\startdata
CWISEP J000006.01$-$704851.2 & $18.28 \pm 0.05$ & VIRCAM &    -    &    -    &    -    &    -    & $18.99 \pm 0.27$ & VIRCAM \\
CWISEP J001146.07$-$471306.8 & $19.22 \pm 0.11$ & VIRCAM & $19.79 \pm 0.34$ & VIRCAM &    -    &    -    & $> 18.35$ & VIRCAM \\
CWISEP J003507.77$-$153233.8 & $19.13 \pm 0.03$ & WIRC &    -    &    -    &    -    &    -    &    -    &    -    \\
CWISEP J004158.35+381811.9 & $19.26 \pm 0.07$ & WIRC & $19.54 \pm 0.22$ & WFCAM &    -    &    -    &    -    &    -    \\
CWISEP J010650.61+225159.1 & $17.41 \pm 0.03$ & WFCAM &    -    &    -    &    -    &    -    &    -    &    -    \\
CWISEP J012748.35$-$631056.1 & $17.82 \pm 0.03$ & VIRCAM &    -    &    -    &    -    &    -    & $18.10 \pm 0.19$ & VIRCAM \\
CWISEP J015613.24+325526.6 & $21.46 \pm 0.30$ & WIRC &    -    &    -    &    -    &    -    &    -    &    -    \\
CWISEP J020103.10+293801.8 & $17.91 \pm 0.03$ & WFCAM & $18.19 \pm 0.06$ & WFCAM & $18.24 \pm 0.12$ & WFCAM &    -    &    -    \\
\enddata
\tablenotetext{a}{CWISEP 0938+0634 appears to be in a sky region with nebulosity based on our Gemini $J$ band image. While this may be contaminating its $J$ band photometry, CWISEP 0938+0634 is nevertheless a real compact source with unambiguously confirmed motion (see Figure \ref{fig:radec_plots}).}
\end{deluxetable*}
\end{longrotatetable}

\begin{deluxetable}{ccc}
\tablecaption{$J$ band follow-up for candidates not motion confirmed. \label{tab:j_follow_up_no_motion}}
\tablehead{
\colhead{CWISEP} & \colhead{$J_{MKO}$} & \colhead{origin} \\
\colhead{designation} & \colhead{(mag)} &
}
\startdata
J021243.55+053147.2 & $22.75 \pm 0.78$ & WIRC \\
J042455.68+000221.4 & $19.66 \pm 0.07$ & WIRC \\
J100854.84+203136.6 & $20.73 \pm 0.39$ & WIRC \\
J121358.13+294237.0 & $18.97 \pm 0.05$ & WIRC \\
J131208.16$-$105231.8 & $20.60 \pm 0.09$ & FLAMINGOS-2 \\
J172104.42+595047.7 & $20.42 \pm 0.13$ & WIRC \\
J175746.31+195112.6 & $> 21.36$\tablenotemark{a} & WIRC \\
J221736.94$-$222647.6 & $> 21.56$\tablenotemark{a}$^{,}$\tablenotemark{b} & FLAMINGOS-2 \\
\enddata
\tablenotetext{a}{Quoted limits are $5\sigma$.}
\tablenotetext{b}{A smudge appears at the location of CWISEP 2217$-$2226, for which we quote a $J$ magnitude lower limit.}
\end{deluxetable}

\startlongtable
\begin{deluxetable*}{lcccccc}
\tablecaption{Derived Properties for Motion-confirmed Targets with \textit{Spitzer} Imaging Available \label{tab:derived_properties}}
\tablehead{
\colhead{Name} & \colhead{SpT} & \colhead{$M_{ch2}$} & \colhead{distance} & \colhead{$T_{eff}$} & \colhead{$H_{ch2}$} & \colhead{$V_{tan}$}  \\
 & \colhead{(phototype)} & \colhead{(mag)} & \colhead{(pc)} & \colhead{(K)} & \colhead{(mag)} & \colhead{(km/s)} 
}
\startdata
CWISEP J000006.01$-$704851.2 & 7.0 & $13.21^{+0.31}_{-0.30}$ & $33.5^{+5.1}_{-4.4}$ & $813^{+85}_{-85}$ &  $18.34^{+0.29}_{-0.26}$& $50^{+10}_{-9}$ \\
CWISEP J000110.81$-$093215.5 & 7.5 & $13.43^{+0.31}_{-0.31}$ & $31.0^{+4.7}_{-4.1}$ & $720^{+85}_{-85}$ &  $18.88^{+0.44}_{-0.36}$& $58^{+14}_{-13}$ \\
CWISEP J001146.07$-$471306.8 & 8.5 & $13.73^{+0.33}_{-0.32}$ & $26.1^{+4.2}_{-3.6}$ & $628^{+88}_{-87}$ &  $18.50^{+0.44}_{-0.36}$& $43^{+10}_{-10}$ \\
CWISEP J003507.77$-$153233.8 & 8.0 & $13.67^{+0.31}_{-0.31}$ & $20.5^{+3.2}_{-2.8}$ & $646^{+85}_{-85}$ &  $18.94^{+0.17}_{-0.16}$& $54^{+9}_{-8}$ \\
CWISEP J004158.35+381811.9 & 8.0 & $13.51^{+0.32}_{-0.32}$ & $32.5^{+5.2}_{-4.5}$ & $695^{+89}_{-89}$ &  $18.43^{+0.41}_{-0.34}$& $46^{+11}_{-10}$ \\
CWISEP J005802.63+723330.3 & 4.5 & $12.43^{+0.24}_{-0.24}$ & $73.2^{+8.6}_{-7.8}$ &     -     &  $20.30^{+0.27}_{-0.24}$& $178^{+30}_{-28}$ \\
CWISEP J010527.69$-$783419.3 & 9.0 & $13.94^{+0.32}_{-0.32}$ & $17.9^{+2.8}_{-2.4}$ & $576^{+85}_{-84}$ &  $18.09^{+0.12}_{-0.12}$& $32^{+5}_{-5}$ \\
CWISEP J010650.61+225159.1 & 6.5 & $13.12^{+0.30}_{-0.30}$ & $25.3^{+3.8}_{-3.3}$ & $858^{+83}_{-83}$ &  $17.69^{+0.25}_{-0.22}$& $39^{+7}_{-7}$ \\
\enddata
\tablecomments{Photometric spectral type values are defined such that SpT = 6 for T6, SpT = 7 for T7,... SpT = 11 for Y1.}
\end{deluxetable*}

\begin{deluxetable*}{ccccc}
\tablecaption{Summary of Motion-confirmed Discoveries Meriting Further Follow-up to Better Assess Y dwarf Candidacy \label{tab:possible_y_dwarfs}}
\tablehead{
\colhead{Name} & \colhead{SpT} & \colhead{ch1$-$ch2} & \colhead{[(ch1$-$ch2)$-$2.4]/$\sigma_{ch1-ch2}$} & \colhead{$J - $ch2} \\
 & \colhead{(phototype)} & \colhead{(mag)} & \colhead{(dimensionless)} & \colhead{(mag)}
}
\startdata
CWISEP J144606.62$-$231717.8 & $\geq 11.0$ & $3.709 \pm 0.435$ & 3.01 & $ > 6.56$ \\
CWISEP J225628.97+400227.3 & $\geq 11.0$ & $3.014 \pm 0.227$ & 2.71 & $5.90 \pm 0.41$ \\
CWISEP J193518.58$-$154620.3 & $\geq 11.0$ & $2.984 \pm 0.034$ & 17.15 & $ > 6.17$ \\
CWISEP J223022.60+254907.5 & $\geq 11.0$ & $2.830 \pm 0.293$ & 1.47 & $ > 4.87$ \\
CWISEP J094005.50+523359.2 & $\geq 11.0$ & $2.766 \pm 0.175$ & 2.09 & $ > 5.50$ \\
CWISEP J023842.60$-$133210.7 & $\geq 11.0$ & $2.729 \pm 0.220$ & 1.50 & $ > 5.13$ \\
CWISEP J040235.55$-$265145.4 & $\geq 11.0$ & $2.720 \pm 0.152$ & 2.10 & $ > 4.89$ \\
CWISEP J235644.78$-$481456.3 & 10.5 & $2.687 \pm 0.220$ & 1.31 & $5.73 \pm 0.28$ \\
\enddata
\tablenotetext{a}{The relatively blue $J - $ch2 color of CWISEP 2355+3804 suggests that it is a late T dwarf despite its \textit{Spitzer}-based phototype of Y0.}
\tablecomments{Photometric spectral type values are defined such that SpT = 6 for T6, SpT = 7 for T7,... SpT = 11 for Y1.}
\end{deluxetable*}

\end{document}